\magnification=1200
\hsize=15.0truecm
\vsize=24truecm
\parindent=1.0truecm
\baselineskip=0.8truecm plus 0.1truecm
\parskip=0.1truecm plus 0.01truecm

\centerline{\bf THE GEOMETRY OF ENTANGLEMENT: METRICS, CONNECTIONS}
\centerline{\bf AND THE GEOMETRIC PHASE}

\bigskip
\bigskip
\centerline{\bf P\'eter  L\'evay }
\bigskip
\centerline {\it Department of Theoretical Physics, Institute of Physics, Technical University of Budapest}
\centerline{\it H-1521 Budapest, Hungary}
\bigskip
\bigskip

\centerline{\bf Abstract}
Using the natural connection 
 equivalent to the $SU(2)$ Yang-Mills instanton
on the quaternionic Hopf fibration of $S^7$ over
the quaternionic projective space ${\bf HP}^1\simeq S^4$ with an $SU(2)\simeq S^3$ fiber
the geometry of entanglement for two qubits is investigated.
The relationship between base and fiber i.e. the twisting of the bundle corresponds to the entanglement of the qubits.
The measure of entanglement can be related to the length
of the shortest geodesic with respect to the Mannoury-Fubini-Study metric
on ${\bf HP}^1$  
between an arbitrary entangled state, and the separable state nearest to it.
Using this result an interpretation of the standard Schmidt decomposition in geometric terms is given.
Schmidt states are the nearest and furthest separable ones lying on, or the ones obtained
by parallel transport along the geodesic passing through the entangled state.
Some examples showing the correspondence between the anolonomy of the connection and entanglement via the geometric phase is shown.
Connections with important notions like the Bures-metric, Uhlmann's connection, the hyperbolic structure for density matrices and anholonomic quantum computation are also pointed out. 

\vfill
\eject
\bigskip
\bigskip
\centerline{\bf I. Introduction}
\bigskip
Since the advent of quantum computation the importance of quantum entanglement 
cannot be overestimated. 
Maximally entangled states 
have made their debut to physics via Bell-type inequalities [1] exemplifying
measurable differences between classical and quantum predictions.
Recently entangled states have become important via their basic use in quantum computation processes [2], teleportation [3], dense coding [4], and quantum key distribution [5].
In the light of such applications it has become evident that quantifying entanglement and understanding  its geometry is a problem of basic importance.
Efforts have been made to quantify entanglement by introducing suitable measures for it [6,7] . These approaches emphasize the difference between entangled and separable states 
by introducing measures usually related to the entropy of the states [8].
At the same time some authors have  pointed out correspondences between the notion of entanglement  and the basic geometry
of the space of states .
The space of states for spin-like systems with the composite Hilbert space  ${\cal H} \simeq {\bf C}^{n}$   is ${\bf CP}^{n-1}$ the $n-1$ dimensional complex projective space. Here the slicing of the space of states for  submanifolds of fixed entanglement 
was introduced and illustrated [9,10].
It was also shown how algebraic geometric ideas can be used to 
study two qubit entanglement within the framework of geometric quantum mechanics [11].
Recently in an interesting paper Mosseri and Dandoloff [12]
used the quaternionic Hopf fibration  to take another look at the problem
of characterizing the geometry of two qubit entanglement.
Their results have been generalized to three qubits by using the next Hopf fibration based on octonions [13].

This paper can be regarded as a further development in understanding entenglement in geometric terms.
For illustrativ purposes we take the simplest two-qubit case
and use the convenient quaternionic representation [12] of two qubit entanglement.
In this picture the two qubit Hilbert space is fibered over the four-sphere  $S^4$ which is isomorphic to ${\bf HP}^1$ the one dimensional quaternionic projective space.
This four-sphere is sliced to submanifolds of fixed entanglement.
Our key idea is the observation that the quaternionic Hopf bundle can be equipped with a natural connection enabling a geometric means for comparing
states belonging to submanifolds of different entanglement.
This connection provides a splitting for vectors corresponding to entangled states
to parts representing their horizontal and vertical components.
Using this splitting a natural metric (the Mannoury-Fubini-Study metric) is induced
on ${\bf HP}^1$ which is essentially the standard metric on the four spehere  $S^4$ expressed in terms of stereographically projected coordinates.
The geodesic distance with respect to this metric provides a natural tool for
quantifying entanglement. This simple picture gives a further understanding of the results
of Brody and Hughston [11] quantifying entanglement by the geodesic distance
between the entangled state in question and the {\it nearest} separable state
with respect to the standard Fubiny-Study metric on ${\bf CP}^3$.
The important new ingredient of our paper is the possibility of using the non-Abelian geometric phase (the anholonomy of the natural connection) in obtaining a further insight to the geometry of two qubit entanglement.
Our approach beeing interesting in its own right also gives an interesting application of the idea of  holonomic quantum computation [14].

The organization of this paper is as follows.
In Section II. we briefly summarize some basic background material concerning
two-qubit entanglement.
In Section III. we reformulate the results of Mosseri et.al. on the Hopf fibering of the two qubit Hilbert space in a formalism convenient for our purposes.
In Section IV. we introduce our geometric structures, the connection and the metric.
In Section V. using this formalism we show that the geodesic distance between an entangled state $\Psi$ and the closest separable state is a convenient quantity characterizing entanglement. 
In fact this quantity is expressed in terms of the concurrence of the entangled state in question. The nearest and furthest separable states obtained by parallel transport with respect to the instanton connection are just the ones appearing in the Schmidt decomposition of $\Psi$. 
 In Section VI.
some examples showing the correspondence between the anholonomy of the natural connection , anholonomic quantum computation and entanglement via the geometric phase is shown.
In Section VII. 
some connections with important notions like the Bures-metric and Uhlmann's connection for density matrices are also given.
We also relate these notions to the hyperbolic structure of the space of
reduced density matrices.
The conclusions and some comments are left for Section VIII.
\bigskip
\bigskip
\centerline{\bf II. Two-qubit entanglement}
\bigskip
In this section we summarize well-known results concerning two-particle pure-state entanglement. Although formulas below are valid for wave functions of both
particles belonging to a finite $N$ dimensional Hilbert space ${\cal H}^{N}$, we have in mind the $N=2$ case i.e. two qubits. 
As a starting point we write the two-particle wave function as

$$\vert \Psi\rangle ={1\over {\sqrt{N}}}\sum_{\alpha,\beta=0}^{N-1}C_{\alpha\beta}\vert \alpha\beta\rangle,
\quad \vert \alpha\beta\rangle\equiv {\vert \alpha\rangle}_1\otimes{\vert \beta\rangle}_2,\eqno(1)$$

\noindent
where ${\vert \alpha\rangle}_{1}$  and ${\vert \beta\rangle}_2$ are orthonormal bases for subsystems $1$ and $2$
and $\langle\Psi\vert\Psi\rangle=1$.
The pure state density matrix of the total system is $\rho=\vert\Psi\rangle\langle \Psi\vert$.
The reduced density matrices  ${\rho}_1$ and ${\rho}_2$ characterizing the state of the system available to an observable capable of performing local manipulations merely on subsytem $1$ respectively on subsystem $2$  
are given by the expressions

$${\rho}_1={\rm Tr}_2{\rho}={1\over N}CC^{\dagger},\quad
  {\rho}_2={\rm Tr}_1{\rho}={1\over N}\overline{C^{\dagger}C},\eqno(2)$$

\noindent
where it is understood that ${\rho}_1$ and ${\rho}_2$ are $N\times N$ matrices expressed in the base ${\vert \alpha\rangle}_1$ and ${\vert \beta\rangle}_2$
respectively.
The von Neumann entropy corresponding to the $i$-th subsystem
is defined as
$$S_{i}=-{\rm Tr} {\rho}_i{\log}_2{\rho}_i=-\sum_{n,{\lambda}_n^{(i)}\neq 0}{\lambda}_n^{(i)}{\log}_2{\lambda}_n^{(i)},\quad i=1,2.\eqno(3)$$

\noindent
Using the fact that $CC^{\dagger}$ and $C^{\dagger}C$ are hermitian matrices having the same real nonzero eigenvalues one can see that
$S_1=S_2$.

For {\it maximally entangled states} we have ${\rho}_1={\rho}_2={1\over N}I$ where 
$I$ is the $N\times N$ identity matrix.
In this case from (2) and (3) it follows that $C\in U(N)$ and $S={\log}_2N$,
hence in particular for two qubits we have $C\in U(2)$ and $S=1$.
For {\it separable states} we have $C_{\alpha\beta}=X_{\alpha}Y_{\beta}$, hence in this case $\vert\Psi\rangle$ can be written in the product form $\vert \psi\rangle\otimes \vert \varphi\rangle$.
One can readily show that $\vert\Psi\rangle$ is separable if and only if
$S=0$.
Since for separable states the matrix $C$ is a dyadic product of two vectors
the partially traced density matrices and $C$ are all of rank one.
For the two-qubit $N=2$ case it means that $\vert \Psi\rangle$ is separable if and only if ${\rm det} C=0$.
Separable and maximally entangled states are extremal in the sense that they
give the minimum and maximum values for the von Neumann entropy. 
In between these cases lie states of intermediate entanglement
characterized by the values $0<S<{\log}_2N$.

It is well-known that an arbitrary state $\vert\Psi\rangle \in {\cal H}^{N}\otimes {\cal H}^{N}$ expressed as in (1) can be transformed to the Schmidt form [15] 

$$\vert\Psi\rangle=\sum_{j=0}^{N-1}\sqrt{{\lambda}_j}{\vert j\rangle}_1\otimes {\vert j\rangle}_2\eqno(4)$$
\noindent
by means of local unitary $U(N)\times U(N)$ transformations acting independently on the two subsystems.
The nonnegative real numbers ${\lambda}_j$ are the eigenvalues of the reduced density matrices, hence they sum to one in accordance with the property 
${\rm Tr}{\rho}_{1,2}=1$.
Notice that in the sum only the nonzero eigenvalues contribute which are the same for both reduced density matrices.
The orthonormal states ${\vert j\rangle}_{1,2}$ can be obtained by finding the eigenvectors corresponding to the nonzero eigenvalues of the reduced density matrices ${\rho}_{1,2}$.

Let us give explicit expressions for the $N=2$ case!
We write

$$\vert\Psi\rangle={1\over {\sqrt{2}}}\left(a\vert 00\rangle +b\vert 01\rangle +c\vert 10\rangle +d\vert 11\rangle\right),\quad {\rm i.e.}\quad
C_{\alpha\beta}=\left(\matrix{a&b\cr c&d\cr}\right)\eqno(5)$$

\noindent
Let us define the complex numbers

$$z\equiv \overline{a}c+\overline{b}d,\qquad w\equiv{ad-bc},\qquad  \zeta \equiv \overline{a}b+\overline{c}d.\eqno(6)$$
Then
the reduced density matrices are

$${\rho}_1={1\over 2}\left(\matrix{{\vert a\vert}^2+{\vert b\vert}^2&
\overline{z}\cr z&{\vert c\vert}^2+{\vert d\vert}^2\cr}\right)\qquad
{\rho}_2={1\over 2}\left(\matrix{{\vert a\vert}^2+{\vert c\vert}^2&\overline{\zeta}\cr
\zeta&{\vert b\vert}^2+{\vert d\vert}^2\cr}\right).\eqno(7)$$

Due to the normalization condition $\langle\Psi\vert \Psi\rangle =1$
we have ${\rm Tr}{\rho}_{1,2}=1$, moreover by virtue of (2)  
${\rm Det}{\rho}_{1,2}={1\over 4}{\vert{\rm Det}C\vert}^2={1\over 4}{\vert w\vert}^2.$
The magnitude of $w$ is the {\it concurrence} satisfying the relation
$0\leq {\cal C}\equiv\vert w\vert\leq 1$ [7].
It is obvious that for separable states one has ${\cal C}=0$.
For maximally entangled states $C\in U(2)\simeq U(1)\times SU(2)$ hence 
${\cal C}=\vert{\rm Det}C\vert=1$.
The eigenvalues of the reduced density matrices in terms of the concurrence read as 

$${\lambda}_{\pm}={1\over 2}\left(1\pm\sqrt{1-{\cal C}^2}\right).\eqno(8)$$

\noindent
Using this result it is reasonable to define the entanglement of a two-qubit pure state $\vert\Psi\rangle$   
to be its von Neumann entropy [16] i.e.

$$E(\Psi)=-{\rm Tr}{\rho}_1{\log}_2{\rho}_1=
-{\rm Tr}{\rho}_2{\log}_2{\rho}_2=-{\lambda}_+{\log}_2{\lambda}_+-
{\lambda}_-{\log}_2{\lambda}_-.\eqno(9)$$

\noindent
Since ${\cal C}$ has the same range for its values 
and is monotonically related to $E(\Psi)$,  the concurrence can be regarded as 
a measure of entanglement in its own right.

Employing {\it local} $U(2)\times U(2)$ rotations our $\vert{\Psi}\rangle$ can be transformed to the (4) Schmidt form i.e. we have

$$\vert\Psi\rangle =\sum_{j,\alpha ,\beta =0,1}\sqrt{{\lambda}_j}{\vert \alpha\rangle}_1U_{\alpha j}\otimes{\vert\beta\rangle}_2V_{\beta j},\quad U^{\dagger}U=V^{\dagger}V=I,
\eqno(10)$$

\noindent
where ${\lambda}_{0,1}={\lambda}_{+,-}$.
Using this expression one can check that (see also [9])

$$ {1\over {\sqrt{2}}}C=UD{V}^{T},\qquad {\rho}_1=UD^2U^{\dagger},\qquad
{\rho}_2=VD^2V^{\dagger},\eqno(11)$$

\noindent
where $D$ is the diagonal matrix containing the square root of the eigenvalues ${\lambda}_{\pm}$ in its diagonal.
Taking the magnitude of the determinant of the expression for $C$ in Eq. (11) shows that  local transformations presereve the concurrence, hence they do not change the degree of entanglement. 
This observation fulfills our expectations that entanglement can be changed only by {\it global} (i.e. $U(4)$ transformations).

\bigskip
\bigskip
\centerline{\bf III. Entanglement and the quaternionic Hopf fibration}

\bigskip
According to Equation (5) the set of normalized states is characterized by
the constraint $\vert a\vert ^2+\vert b\vert ^2+\vert c\vert ^2+\vert d \vert ^2=2$ (recall our convention of pulling out a factor of ${1\over {\sqrt{2}}}$
from the expansion coefficients of $\vert \Psi\rangle $), is the seven-sphere $S^7$.
The basic observation of [12] is that for understanding the geometry of two-qubit entanglement it is useful to fibre $S^7$ over the four dimensional sphere $S^4$ by employing the second Hopf-fibration.
Moreover, it is convenient to introduce quaternionic notation for our basic quantities 
since the geometry of this fibration then easily described.
Let us represent an element of $ S^7$ by the
quaternionic spinor, i.e. let

$$ {1\over {\sqrt{2}}}\left(\matrix{a\cr b\cr c\cr d}\right)\mapsto \left(\matrix{u_0\cr u_1\cr
}\right)\equiv
{1\over {\sqrt{2}}}\left(\matrix{a +b {\bf j}\cr c+d{\bf j}\cr}\right).\eqno(12)$$

\noindent
The quaternionic units ${\bf i}$, ${\bf j}$ and ${\bf k}$ with squares equal to $-1$ satisfy the usual relations  ${\bf ij}=-{\bf ji}= {\bf k}$ plus
similar ones obtained  by employing cyclic permutations of the symbols ${\bf ijk}$. In this way an arbitrary quaternion $q\in {\bf H}$  can be expressed as a pair of complex numbers 

$$q= q_1+q_2{\bf i}+q_3 {\bf j}+q_4{\bf k}=(q_1+q_2{\bf i})+(q_3+q_4 {\bf i}){\bf j},\eqno(13)$$
\noindent
where the components $q_l,\quad l=1,2,3,4$ are real numbers.
The conjugate quaternion $\overline{q}$ is obtained by changing the signs in front of the terms containing ${\bf i, j}$ and ${\bf k}$ 

$$\overline{q}= q_0-q_1{\bf i}-q_2 {\bf j}-q_3{\bf k}=(q_0-q_1{\bf i})-(q_2+q_3 {\bf i}){\bf j}.\eqno(14)$$
\noindent
Recall also that due to the noncommutativity of quaternionic multiplication we have $\overline{qp}=(\overline{p})(\overline{q})$, and the norm squared of a quaternion $q$ is defined as  the real number $\vert q\vert ^2=\overline{q}q$.
Now we are ready to define the second Hopf fibration by the map $\pi$ as follows

$$ \pi: S^7\to S^4,\quad 
\left(\matrix{u_0\cr u_1\cr}\right)\mapsto
\left( 2u_1\overline{u_0},
 {\vert u_0\vert}^2-{\vert u_1\vert}^2\right)\equiv (\xi,{\xi}_0),\eqno(15)$$

\noindent
where 
$$ {\xi}=({\xi}_1+{\xi}_2{\bf i})+({\xi}_3+{\xi}_4{\bf i}){\bf j},\quad
{\xi}_{\mu}{\xi}_{\mu}={\xi}_0^2+\overline{\xi}\xi=1,\quad \mu =0,1,\dots 4
\eqno(16)$$

\noindent
 are Cartesian coordinates for $S^4$ (summation for repeated indices is understood). 
Since 
$2u_1\overline{u}_0=(c+d{\bf j})\overline{(a +b{\bf j})}=
(c+d{\bf j})(\overline{a}-{\bf j}\overline{b})=
z+w{\bf j}$
the Cartesian coordinates for $S^4$ can be expressed in terms of the quantities 
defined in (6) 
characterizing two-qubit entanglement, i.e. we have 

$$ {\xi}_0=\pm\sqrt{1-\vert z\vert ^2-\vert w\vert^2},\qquad
{\xi}_1+{\xi}_2{\bf i}=z,\qquad {\xi}_3+{\xi}_4{\bf i}=w.\eqno(17)$$

\noindent
The basic result of [12] was that the mapping $\pi$ is entanglement sensitive.
In this formalism this result is easily reproduced by noticing
that
submanifolds of fixed entanglement are characterized by fixed values for the concurrence
${\cal C}\equiv \sqrt{\vert {\xi}_3\vert ^2+\vert {\xi}_4\vert ^2}$.
Hence separable states are mapped to points of $S^4$ with 
vanishing values for the coordinates ${\xi}_3$ and ${\xi}_4$, i.e they are on the two-sphere  $S^2\subset S^4$  described by the constraint ${\xi}_0^2+{\xi}_1^2+{\xi}_2^2=1$.         
For maximally entangled states we have ${\cal C}=1$, then for these states
we have ${\xi}_0={\xi}_1={\xi}_2=0$. These states are parametrized by a great circle of the "equator" (which is a three-sphere $S^3$) of $S^4$.

It is clear from Eq. (15) that multiplication of the quaternionic spinor from the ${\it right}$ by a {\it unit quaternion} $q$ (i.e. a quaternion with unit norm $\vert q\vert ^2\equiv \overline{q}q=1$) leaves the coordinates ${\xi}_{\mu}$ invariant. Since quaternions of unit norm ("quaternionic phases") form the group $Sp(1)\simeq SU(2)$ it means that entangled states 
related by an $SU(2)$ rotation project to states of the same concurrence.
This gauge degree of freedom corresponds to the fiber of the second Hopf fibration. 
Since $SU(2)\simeq S^3$ the fibration $\pi$ is a one with total space $S^7$,
base space $S^4$, and fiber $S^3$.
The important result of the present paper is that the local gauge transformations of the Hopf fibration associated with this $Sp(1)\simeq SU(2)$ degree of freedom give rise to a geometric interpretation of 
{\it local} transformations in the {\it second} subsystem not changing the entanglement properties of our two-qubit system. 
The information available for the observer of the {\it first} subsystem
is parametrized by the base space  $S^4$ of the fibration as can be seen from the (7)
form of the reduced density matrix ${\rho}_1$.
By exchanging the parameters $b$ and $c$ in the (12) definition we obtain another representation of the Hopf fibration with Cartesian coordinates ${\eta}_{\mu}$ for the corresponding four-sphere $S^4$. 
The assignment in this case reads ${\eta}={\zeta}+{w}{\bf j}$ with ${\zeta}$ defined by Eq. (6).
In this case as was explained in Ref. [12] the roles of the two qubits are exchanged. Now the local gauge degree of freedom associated with our ignorance of the details of the {\it first} subsystem is represented by the fiber degree of freedom.  The base space parametrizes the reduced density matrix ${\rho}_2$.

The relationship between base and fiber (i.e. the twisting of the bundle)
is just the entanglement of the two qubits.
A natural way of describing this twisting is via the means of introducing a connection on our bundle.
Luckily for the second Hopf fibration we have a canonical connection
the properties of which has been described in many places (see e.g. [17,18]).
This connection is equivalent to the instanton connection well-known to physicists. Moreover, it can be related to a metric on ${\bf HP}^1\simeq S^4$  which is the quaternionic counterpart of the complex Fubini-Study metric on ${\bf CP}^1\simeq S^2$.
Our next task is to describe these quantities, and relate them to our basic ones of Section II. describing the phenomenon of two-qubit entanglement.

\bigskip
\bigskip
\centerline{\bf IV. Sections, connections, and metrics}
\bigskip

First we introduce for two quaternionic spinors $\vert v\rangle$ and $\vert u\rangle$ the scalar product $\langle v\vert u\rangle \equiv\overline{v}^{\alpha}u_{\alpha}=\overline{v}_0u_0+\overline{v}_1u_1$.
Notice that {\it right} multiplication of our spinors with the nonzero quaternion $q$
yields the expression $\langle vq\vert uq\rangle =\overline{q}\langle v\vert u\rangle q$. The vector space ${\bf H}^2$
of quaternionic spinors with this scalar product is an example of a {\it quaternionic Hilbert space} (see Ref. 18. and references therein).
States in quaternionic quantum mechanics based on the space ${\bf H}^N$ are represented by points of the quaternionic space of rays which is just ${\bf HP}^{N-1}$ the $N-1$ dimensional quaternionic projective space.
It is amusing to see that the $N=2$ case of interest for us
yields the space ${\bf HP}^1\simeq S^4$, which is the quaternionic analogue
of the usual Bloch-sphere representation of {\it complex} spinors, i.e.
we have ${\bf CP}^1\simeq S^2$.
The complex Bloch-sphere is of basic importance for the geometric description of complex superposition, likewise the "quaternionic Bloch-sphere" plays a similar role for the geometrical description of quantum entanglement.
Though this correspondence between entanglement and quaternionic quantum mechanics is an interesting idea to follow in its own right, here we work out merely the simplest $N=2$ case and regard the quaternionic Hilbert space formalism merely as a comfortable representation. 

What is interesting for us is that two important geometric quantities can be defined on the space of normalized quaternionic spinors $S^7$  which pull back naturally
to the base space ${\bf HP}^1\simeq S^4$, the metric and the connection.
The first of these is related to the transition probability
${\vert \langle v\vert u\rangle\vert}^2$ and the second to the transition amplitude $\langle v\vert u\rangle$ [18].

Indeed an  invariant distance

$${\cos}^2{{\Delta}_{vu}\over 2}={\vert\langle v\vert u\rangle\vert}^2\quad 0<\Delta<\pi\eqno(18)$$

\noindent
between two not identical, nonorthogonal quaternionic spinors representing entangled states can be defined. It is related to the distance along  
the geodesic connecting the points $\pi(\vert v\rangle)$ and
$\pi(\vert u\rangle)$, representing the corresponding states in ${\bf HP}^1$,
 with respect to the metric which is the obvious quaternionic
generalization of the well-known Fubini-Study metric. By using local coordinates $x_k, k=1,2,3,4$ on an open set ${\cal U}\subset{\bf HP}^1$ parametrizing our spinor $\vert u\rangle$ and putting $dl=\Delta$ it is defined 
by the relation

$$dl^2=g_{km
}dx^k\otimes dx^m=4\left(1-{\vert \langle u+du\vert u\rangle\vert}^2\right).\eqno(19)$$
\noindent
Here it is understood that this equality is valid only up to terms higher than second order in the change of local coordinates.

Moreover, since our entangled states are defined up to right multiplication with a unit quaternion it would be desirable to define a means for comparing
the "quaternionic phases" of states  
of different entanglement.
For the rule of comparing "phases" we adopt the definition that, two such states are "in phase" if $\langle u+du\vert u\rangle =1$ up to second order terms in $du$ (the quaternionic analogue of the so called Pancharatnam connection [19] ).
 By introducing the quantity

$$\Gamma =1-\langle u+du\vert u\rangle\eqno(20)$$

\noindent
to be used later this rule can be restated as $\Gamma =0$ up to a second order term.

In order to enable an explicit construction we have to chose a section for our
bundle. This means that we have to adopt a choice for $\vert u\rangle\in S^7$ parametrized by points of $S^4$. 
If the bundle is nontrivial the best we can do is to chose {\it local sections}.
We chose the section

$$\vert u\rangle =\left(\matrix{u_0\cr u_1\cr}\right)=
{1\over {\sqrt{1+\vert x\vert ^2}}}\left(\matrix{ 1\cr x\cr}\right)q,\eqno(21)$$

\noindent
with $\overline{q}q=1$. In this parametrization we have $x=u_1(u_0)^{-1}$ hence it is valid on the coordinate patch ${\cal U}$ characterized by the constraint $u_0\neq 0$. 
Using this section we can pull back the (19) metric and (20) connection 
to the base space ${\bf HP}^1$ yielding the formulae [18]

$$dl^2={ 4{d\overline{x} dx} \over {(1+\vert x\vert ^2)^2}},\eqno(22)$$

$$\Gamma =\overline{q}\left({{{\rm Im}\overline{x}dx}\over {1+\vert x\vert ^2}}\right)q +\overline{q}dq.\eqno(23)$$

\noindent
Here ${\rm Im}q={1\over 2}(q-\overline{q})$ is the imaginary part of an arbitrary  quaternion $q$. (Similarly the real part of $q$ is defined by ${\rm Re}q={1\over 2}(q+\overline{q})$.)
The quantity

$$A={\rm Im}{\overline{x}dx\over {1+\vert x\vert ^2}} \eqno(24)$$

\noindent
is an $sp(1)\simeq su(2)$-valued one-form (non-Abelian gauge-field) equivalent to the standard $SU(2)$ instanton with self-dual curvature and second Chern-number $C_2=1$ [17-18]. 
Notice that according to Eq. (20), Pancharatnam connection ($\Gamma =0$) yields
a condition for parallel translation of quaternionic phases.
Indeed, using Eq. (23) 
with a suitable boundary condition we obtain the usual differential equation  
of parallel transport.
For a curve ${\cal C}$ lying entirely in ${\cal U}$  with initial and end points  being $q(0)=1$ and $q(\tau)$, we obtain the standard path ordered solution

$$q(\tau)={\bf P}\exp \left( -\int_{\cal C}A\right). \eqno(25)$$

Observe that our coordinates $x=x_1+x_2{\bf i}+x_3{\bf j}+x_4{\bf k}$
 used in the (21) section
are related to the Cartesian coordinates ${\xi}={\xi}_1+{\xi}_2{\bf i}+{\xi}_2{\bf j}+{\xi}_4{\bf k}$ and ${\xi}_0$ as

$$\xi ={2x\over {1+\vert x\vert ^2}},\qquad {\xi}_0={1-\vert x\vert ^2\over {1+\vert x\vert ^2}}.\eqno(26)$$

\noindent
Indeed,  the coordinates $(x_1,x_2,x_3, x_4)$  are obtained from stereographically projecting the sphere $S^4$ from its south pole to ${\bf R}^4\cup\{\infty\}$. 
It is straightforward to check in these coordinates that 
$dl^2= d{\xi}_0^2+d{\xi}_1^2+d{\xi}_2^2+d{\xi}_3^2+d{\xi}_4^2$ (the standard line element on $S^4$) is just (22). 
Let us define $R^2={\xi}_0^2+{\xi}_1^2+{\xi}_2^2$, then we have the relation 
$0\leq {\cal C}^2=1-R^2\leq 1$ where ${\cal C}$ is the concurrence. Using $R$ and the polar coordinates $(R,\Theta, \Phi)$
on the unit ball ${\bf B}^3$ originally parametrized by the coordinates $({\xi}_0,{\xi}_1, {\xi}_2)$
we have

$$d{\xi}_0^2+d{\xi}_1^2+{\xi}_3^2=dR^2+R^2d{\Omega},\quad {\rm where}\quad
d{\Omega}=d{\Theta}^2+{\sin\Theta}^2 d{\Phi}^2.\eqno(27)$$

\noindent
Hence in terms of the concurrence ${\cal C}$  (or alternatively $R$) and these polar coordinates we have for the line element on the base space $S^4$ 
the expression 

$$dl^2={d{\cal C}^2\over {1-{\cal C}^2}}+{\cal C}^2d{\chi}^2+(1-{\cal C}^2)d{\Omega}={dR^2\over {1-R^2}}+(1-R^2)d{\chi}^2+R^2d{\Omega}^2, \eqno(28)$$

\noindent
where $w=\vert w\vert e^{i\arg w}={\cal C}e^{i\chi}$.
For separable states we have ${\cal C}=0$, hence this line element is reduced to $d\Omega$ the one for the two-sphere $S^2$ corresponding to one of our separate qubits in the base.  
For maximally entangled states we have $R=0$ then the line element characterizing our base qubit reduces to the one of a circle i.e. $dl^2=d{\chi}^2$.
The other qubit in all cases is associated with the fiber of unit quaternions.
The relationship between the two qubits associated with the base and fiber in all cases can be described by parallel transport with respect to the connection {$\Gamma$ .
To gain some insight into this relationship as a first step we have to express the pull-back one form $A$
in terms of our complex coordinates $z$ and $w$.

For this we combine Eqs. (17) and (26) to see that the quaternionic phase of $x$ (i.e. the unit quaternion $p\equiv {x\over  {\vert x\vert}}$ ) can be expressed as

$$p\equiv {x\over {\vert x\vert}}={z+w{\bf j}\over {\sqrt {\vert z\vert ^2 +\vert w\vert ^2}}}\in Sp(1)\simeq SU(2).\eqno(29)$$

\noindent
Moreover, since the sum of the squared magnitudes of $\xi$ and ${\xi}_0$ equals one it is useful to represent them as 

$$\sin\theta ={2\vert x\vert \over {1+\vert x\vert ^2}},\quad \cos\theta
={1-\vert x\vert ^2\over {1+\vert x\vert ^2}}\eqno(30)$$

\noindent
with $0\leq \theta <\pi$
The parametrization  $\xi=\sin\theta p$, ${\xi}_0=\cos\theta$
with $p\in S^3$ corresponds to introducing polar coordinates for $S^4$. 
However unlike for the usual parametrization we favour $z$ and $w$ more than $\theta$ hence we express it in terms of these quantities as
$\theta =\arcsin(\sqrt{\vert z\vert ^2+\vert w\vert ^2})$.
In this parametrization containing quantities characterizing the entanglement properties of our qubits the section of Eq. (21) reads as

$$\vert u\rangle =\left(\matrix{\cos{\theta \over 2}\cr \sin{\theta\over 2}p\cr}\right)q={1\over {\sqrt 2}}\left(\matrix{\sqrt{1\pm\sqrt{1-\vert z\vert ^2-\vert w\vert ^2}}\cr
\sqrt{1\mp\sqrt{1-\vert z\vert ^2-\vert w\vert ^2}}{z+w{\bf j}\over {\sqrt{\vert z\vert ^2+\vert w\vert ^2}}}\cr}\right)q,\eqno(31)$$

\noindent
where $\pm$ ($\mp$) corresponds to sections over the northern or southern hemispheres.
Since $p,q\in Sp(1)$ i.e. they are quaternionic phases, the parametrization
in terms of $\theta$ and the pair  $(p,q)$ is of the same form as the well-known parametrization of a complex spinor associated with the Bloch-sphere.  
However, the second equality also shows the meaning of these parameters in terms of the entanglement parameters.  
Comparing Eq. (31) with Eq. (12) we realize that on the open set ${\cal U}$ we can always chose a section for which our parameter $b$ equals zero. 
For later use here we also remark that in this ($b=0$) parametrization a formula between our complex parameters $z$, $w$ and $\zeta$
of Eq. (6) holds  

$$\zeta ={1\over 2}\left(1\mp\sqrt{1-\vert z\vert ^2-\vert w\vert ^2}\right){w/z\over {1+\vert w/z\vert ^2}}\equiv {\sin}^2{\theta /2}{r\over {1+\vert r\vert ^2}}, \quad r\equiv w/z.\eqno(32)$$

\noindent
Expressing our (24) instanton gauge-potential in terms of the complex coordinates
$z$ and $w$ we obtain on ${\cal U}$ the expression

$$A={1\over 2}(1-\cos \theta){\rm Im}(\overline{p}dp) = 
{{\rm Im}\left(\overline{z}dz+\overline{w}dw+(\overline{z}dw-wd\overline{z}){\bf j}\right)
\over {2\sqrt{1\pm\sqrt{1-\vert z\vert ^2-\vert w\vert ^2}}}}.
\eqno(33)$$

\noindent
We note that for another coordinate patch ${\cal V}$ with $u_1\neq 0$
we would obtain an $Sp(1)$  gauge-transformed expression for $A$ [18].
From Eq. (33) we see that for separable states ($w=0$) $A$
 defined on the submanifold $S^2$ (the boundary of the unit ball ${\bf B}^3$) of $S^4$ has the form

$$A={1\over 2}(1-\cos{\Theta})d{\Phi},\quad {\rm where}\quad {\Phi}\equiv \arg{z},\quad \cos\Theta =\pm
\sqrt{1-\vert z\vert ^2}\eqno(34)$$

\noindent
which is the $U(1)$ gauge potential of a magnetic monopole with pole strength ${1\over 2}$.
Hence we see that when moving entirely in the $S^2\simeq{\partial}{\bf B}^3$ submanifold of separable states the relationship between the qubit in the base and the one in the fiber is  characterized  merely by the possible occurrence of a $U(1)$  anholonomy factor.
For maximally entangled states we have $\vert w\vert =1$  i.e. $w=e^{i\chi}$ and $\vert z\vert =0$, hence in this case we have the one-form 
$A=-{1\over 2}{\rm Im}\left(dw\over w\right)=-{1\over 2}d\chi$ living on the great circle of the equator $S^3$ of $S^4$.
States parametrized by the points of this circle belong to  the $Sp(1)\simeq S^3$ fiber.
Parallel transporting an element $q\in Sp(1)$ along this circle with respect to this one-form $A$ yields an anholonomy factor of $-1$ or $+1$ depending on the winding number of traversals beeing even or odd.
In this way we have obtained an alternative proof for the well-known fact that
the manifold of maximally entangled states in ${\bf CP}^3$ is $Sp(1)/{\bf Z}_2\simeq S^3/{\bf Z}_2$. ($S^7$ is also fibered over ${\bf CP}^3$ with the $U(1)\simeq S^1$ fiber corresponds now to our circle parametrized by the angle $\chi$.)
For states of intermediate entanglement labelled by the values of $\theta$ in the interval $0<\theta <{\pi\over 2}$
from Eq. (31) we have a mixing  between the complex coordinates $z$ and $w$.
This will result in a more complicated pattern for the anholonomy properties, reflecting the richness of the entanglement possibilities for the qubits. 
An explicit example for this phenomenon will be given in Section VI.  

\bigskip
\bigskip
\centerline{\bf V. The geometrical meaning of the Schmidt decomposition}
\bigskip

According to Eqs. (22) and (23) a line element and the pull-back of a connection  can be induced on our space ${\bf HP}^1\simeq S^4$ which can be sliced to submanifolds of fixed entanglement.
Now we make use of these facts to give geometrical interpretation to the Schmidt decomposition for two qubits.
In order to do this we have to characterize a special subclass of geodesics in $S^7$ that project to geodesics on $S^4$.
For this we consider a curve $C=\vert u(s)\rangle\subset S^7$.
Using Eq. (19) for this curve we have $dl^2=4\left(1-\vert \langle u(s+ds)\vert u(s)\rangle\vert ^2\right)$ up to terms second order in $ds$.
Taylor expanding this expression a formula for $dl^2$ is obtained

$$dl^2=4\vert \vert Q(s)\dot u(s)\vert\vert ^2 ds^2,\quad 
 \quad Q(s)=I-P(s),\quad
P(s)=\vert u(s)\rangle\langle u(s)\vert, \eqno(35)$$

\noindent
where $\vert\dot u(s)\rangle\equiv{d\over {ds}}\vert u(s)\rangle$.
Right multiplication by a unit quaternion $q(s)$ leaves invariant the projector $Q(s)$,
but $\vert\dot u(s)\rangle$ transforms to 
$\vert{\dot u(s)}\rangle q(s)+\vert u(s)\rangle \dot{q}(s)$.
However, since $Q(s)\vert u(s)\rangle =0$ we see that $dl^2$ is gauge invariant hence it can be used to define the length of the "shadow" curve $\pi(C)$ in $S^4$ of an arbitrary curve $C\subset S^7$. 
Moreover, notice that expression (35) is also reparametrization invariant.
Now we can characterize geodesics in $S^4$ 
in the following way.
Geodesics in $S^4$ are those curves $\pi(C)$ through $\pi(\vert u(s_1)\rangle )$ and
 $\pi(\vert u(s_2)\rangle)$
for which the following
reparametrization and gauge invariant functional

$$L[C]=2\int_{s_1}^{s_2}ds\vert\vert Q(s)\dot u(s)\vert\vert,\quad C\subset S^7\eqno(36)$$

\noindent
is stationary.
The variation of a similar functional for the complex case
and the derivation of the geodesics was already given in Ref. [20].
For the quaternionic case the same steps has to be taken with the important difference that quaternions do not commute so we have to be careful in grouping terms. However, since the variation $\delta L[C] $ is a real number it can be represented
as the integral of the real part of a quaternion depending on $s$.
Luckily we can cyclically permute the quaternionic entries under 
the operation of taking the real part (it is just the operation of taking the trace when we interpret the quaternions as two-by-two matrices) so the derivation is a straightforward excercise of following the steps described on pages  of 220-223 of Ref. [20].
The result is the following.
Using gauge invariance we can find a solution $\vert u(s)\rangle$ with initial vector $\vert u_i\rangle =\vert u(0)\rangle $ to the geodesic equation on $S^7$ 
which is {\it horizontal} i.e. parallel transported along the geodesic
$\pi(\vert u(s)\rangle )$  with initial point $\pi(\vert u_i\rangle)$ in $S^4$.
Moreover, exploiting the reparametrization invariance it is {\it affinely parametrized} i.e. $\vert\vert \dot u(s)\vert\vert$ is constant along $C\subset S^7$. Such affine parametrizations are unique up to linear inhomogeneous changes in s, for convenience we chose the parametrization for which $\vert\vert\dot u(s)\rangle\vert\vert={1\over 4}$.
In particular a geodesic $C$ starting from $\vert v\rangle$ which is the horizontal lift of the shadow geodesic $\pi(C)$ connecting $\pi(\vert v\rangle)$ and some other point 
$\pi(\vert u\rangle)$ is of the form

$$\vert u(s)\rangle =\cos{s\over 2}\vert{\phi}_1\rangle +\sin {s\over 2}
\vert{\phi}_2\rangle,\qquad \langle {\phi}_i\vert{\phi}_j\rangle ={\delta}_{ij},
\quad i,j=1,2\eqno(37)$$

\noindent
where

$$\vert{\phi}_1\rangle =\vert v\rangle,\quad
  \vert{\phi}_2\rangle =\left(\vert u^{\prime}\rangle - \cos{\Delta\over 2}\vert v\rangle\right)/\sin{\Delta\over 2},\eqno(38)$$

\noindent
and

$$\vert u^{\prime}\rangle =\vert u\rangle{\langle u\vert v\rangle
\over { \vert\langle u\vert v\rangle\vert}}, \quad
\vert\langle u\vert v\rangle\vert =\cos{\Delta\over 2}.\eqno(39)$$

\noindent
Notice that in Eq. (39) in accordance with our conventions the quaternionic phase multiply the state $\vert u\rangle$ from the {\it right}.
It is now understood that as was claimed in Eq. (18) $\Delta$ is the geodesic distance between
the points $\pi(\vert v\rangle $ and
 $\pi(\vert u\rangle $.

Having the geodesic distance at our disposal, let us now define the measure of entanglement as the distance between an arbitrary entangled state and the {\it  separable state nearest to it}. This idea has already been proposed and illustrated in [11] for the case of two qubits  
using algebraic geometric methods on the state space ${\bf CP}^3$.
This space can be regarded as the base for an abelian $U(1)$ fibration of $S^7$ hence it gives rise to an alternative parametrization for submanifolds of fixed entanglement. 
However in contrast to [11], when using instead the Hopf fibration of $S^7$  
the fiber is the non-Abelian group $Sp(1)$ of quaternionic phases making it possible for the two qubits to reside in different spaces, the base and the fiber respectively. The relationship between the qubits, i.e. their entanglement
is measured by the twisting of the bundle.
Hence it is instructive to see by giving an alternative proof, how naturally this measure of entanglement is encoded into the structure of the Hopf bundle.

In order to see this first we chose a representative $\vert u\rangle$ of our entangled state in the (31)  
form with $q=1$ and for the
unknown separable state 
in the similar 

$$\vert v\rangle =\left(\matrix{\cos{\sigma\over 2}\cr {\sin{\sigma\over 2}}e^{i\varphi}\cr}\right)\eqno(40)$$

\noindent
form.
This representative has $b=d=0$ hence ${\cal C}=0$, moreover it is already of the form of our standard section valid on ${\cal U}$.
We have to find the nearest separable state to $\vert u\rangle$, meaning we have to determine
$\sigma$ and $\varphi$ as a function of the entanglement coordinates $z$ and $w$.
In order to do this we have to maximize the expression
$\cos^2{\Delta}_{vu}/2=\vert\langle v\vert u\rangle\vert ^2$.
A short calculation yields for this quantity the expression

$$\vert\langle v\vert u\rangle\vert ^2={\cos}^2{\Delta\over 2}={\cos}^2{\sigma\over 2}{\cos}^2{\theta\over 2}+{\sin}^2{\sigma\over 2}{\sin}^2{\theta\over 2}+{1\over 2}\sin{\sigma}{\rm Re}(
e^{-i\varphi}(z+w{\bf j})).\eqno(41)$$

\noindent
	Since our sections are merely local ones living on ${\cal U}$ we should exclude the south pole ($\sigma=\theta =\pi$) hence we have  $0<\theta < \pi$ and $0 <\sigma <\pi$. For these values (41) is maximal
if $\varphi =\arg z$ .
In this case we are left with the expression
${\cos}^2{\sigma
\over 2}{\cos}^2{\theta\over 2}+{\sin}^2{\sigma\over 2}{\sin}^2{\theta\over 2}+{1\over 2}\vert z\vert\sin\sigma$
to be maximized with respect to changes in $\sigma$.
As one can check this quantity is maximal provided $\tan \sigma ={\vert z\vert\over {\cos\theta}}$. Hence separable states $\vert v\rangle$ nearest to our entangled state $\vert u\rangle$ labelled by the complex numbers $z$ and $w$ are characterized by the angles

$$\cos{\sigma}=\pm\sqrt{1-{\vert z\vert ^2\over {1-\vert w\vert ^2}}},\quad \varphi=\arg z.\eqno(42)$$

\noindent
Using these angles in Eq. (41) we obtain for the distance ${\Delta}_{uv}$
the important formula

$${\cos}^2{{\Delta}_{uv}\over 2}={1\over 2}\left(1+\sqrt{1-{\cal C}^2}\right).\eqno(43)$$

\noindent
Comparing our result with Eq. (8) we see that the value appearing in (43)
is precisely the eigenvalue ${\lambda}_+$ of the reduced density matrix also appearing in the
Schmidt decomposition.
Moreover, it is easy to 
see that the distance of our $\vert u\rangle$ from the state $\vert v^{\prime}\rangle$ orthogonal to 
$\vert v\rangle$ (this state is antipodal to $\vert v\rangle$ in $S^4$)
is related to the other eigenvalue ${\lambda}_-$.
Hence the (9) von-Neumann entropy as a measure of entanglement is
just a special combination of lengths for the shorter and the longer segments of the geodesic linking 
our entangled state $\pi(\vert u\rangle)$ to the  surface $S^2\subset S^4$
of separable states.

What is the meaning of the separable state $\vert v\rangle$ nearest to the entangled state $\vert u\rangle$ ?
It is just the quaternionic representative of one of the states ${\vert j\rangle}_1$ ($j=0,1$) in the (10) Schmidt decomposition.
In order to see this we have to diagonalize the reduced density matrices
${\rho}_1$ and ${\rho}_2$ of Eq. (7).
First we write these matrices in the form

$${\rho}_1={1\over 2}\left(\matrix{1+ {\xi}_0&{\xi}_1-i{\xi}_2\cr {\xi}_1+i{\xi}_2& 1-{\xi}_0\cr}\right)\qquad
{\rho}_2={1\over 2}\left(\matrix{1+{\eta}_0&{\eta}_1-i{\eta}_2\cr
{\eta}_1+i{\eta}_2&1-{\eta}_0\cr}\right), \eqno(44)$$

\noindent
where the coordinates ${\xi}_{\mu}$ are defined in Eq. (17), similarly
 ${\eta}_{\mu}$ is defined by the other Hopf fibration with the roles of $b$ and $c$ in (12) exchanged.
Note also that from the five components of these vectors only the first three
is used. 
We introduce the notation for these vectors

$${\bf v}=\left(\matrix{{\xi}_0\cr{\xi}_1\cr{\xi}_2\cr}\right)=\vert{\bf v}\vert\left(\matrix{\cos\sigma\cr\sin{\sigma}\cos\varphi\cr\sin{\sigma}\sin\varphi\cr}\right),\qquad
{\bf t}=\left(\matrix{{\eta}_0\cr{\eta}_1\cr{\eta}_2\cr}\right)
=\vert{\bf t}\vert\left(\matrix{\cos\tau\cr\sin\tau\cos\epsilon\cr\sin\tau\sin\epsilon\cr}\right)
,\eqno(45)$$

\noindent
where $\sigma$ and $\varphi$ turn out to be precisely the quantities of (42), and the ones $\tau$ and $\epsilon$ are defined by replacing in (42) the coordinate $z$ by $\zeta$ of Eq. (6).
In order to see this notice that
${\bf v}$ and ${\bf t}$ are elements of the unit ball ${\bf B}^3$ i.e. we have
$\vert {\bf v}\vert=\vert{\bf t}\vert=\sqrt{1-\vert w\vert ^2}\leq 1$, and 
${\xi}_0=\cos\theta =\pm\sqrt{1-\vert z\vert ^2-\vert w\vert ^2}$.
In order to diagonalize the reduced density matrices we have to diagonalize
${\bf v \sigma}$ and ${\bf t\sigma}$.
It is well-known that there is no global only local diagonalization of this problem
over ${\bf B}^3$. The reason for this is the fact that the eigenstates of these operators form 
a nontrivial fibration over ${\bf B}^3$ related to the first (complex) Hopf fibration.
Local sections are well-known from studies concerning the geometric phase hence we merely refer to the result [19]. Eigensections of ${\rho}_1$ and ${\rho}_2$ corresponding to the eigenvalue ${\lambda}_+$ of (8) that are singular
on the $-{\xi}_0$ and $-{\eta}_0$ axis are of the form

$$\vert{\bf v}\rangle =\left(\matrix{\cos{\sigma\over 2}\cr \sin{\sigma\over 2}
e^{i\varphi}\cr}\right)\qquad
\vert{\bf t}\rangle =\left(\matrix{\cos{\tau\over 2}\cr \sin{\tau\over 2}
e^{i\epsilon}\cr}\right).\eqno(46)$$

\noindent
For the eigensections belonging to the eigenvalue ${\lambda}_-$ we have

$$\vert{\bf v^{\perp}}\rangle =\left(\matrix{-\sin{\sigma\over 2}e^{-i\varphi}\cr \cos{\sigma\over 2}
\cr}\right)\qquad
\vert{\bf t^{\perp}}\rangle =\left(\matrix{-\sin{\tau\over 2}e^{-i\varphi}\cr \cos{\tau\over 2}
\cr}\right),\eqno(47)$$

\noindent
Looking at Eq. (46) we immediately see that the first vector  $\vert{\bf v}\rangle$ regarded as a quaternion is precisely $\vert v\rangle$ nearest to $\vert u\rangle$.

Now the question arises: what is the meaning of $\vert{\bf t}\rangle$ the Schmidt vector representing one of the orthonormal vectors ${\vert j\rangle}_2$ ($j=0,1$)  used for the other 
qubit?
In the quaternionic notation we know that this vector resides in the fiber of
the Hopf fibration, hence we shoud be able to recover it from the holonomy
of our connection.
In the following we will show that the representative of $\vert{\bf t}\rangle$
is the unit quaternion $Q$ obtained by parallel transporting
our entangled vector  $\vert u\rangle\in S^7$ with respect to the instanton connection along the geodesic segment between $\pi(\vert u\rangle)$ and $\pi(\vert v\rangle)$
representing the closest separable state.

In order to prove this first we represent our entangled state in the (31) form with $q=1$, hence its matrix ${C\over {\sqrt{2}}}$  and the diagonal matrix $D$ of Eq. (11) has the following form

$${C\over {\sqrt{2}}}=\left(\matrix{ \cos{\theta\over 2}&0\cr
\sin{\theta\over 2}{z\over{\sqrt{\vert z\vert ^2+\vert w\vert ^2}}}&
\sin{\theta\over 2}{w\over{\sqrt{\vert z\vert ^2+\vert w\vert ^2}}}\cr}\right),
\quad
D=\left(\matrix{\cos{\Delta\over 2}&0\cr 0&\sin{\Delta\over 2}\cr}\right).
\eqno(48)$$

\noindent
Likewise using (46) and (47) the unitary matrices $U$ and $V$ diagonalizing on ${\cal U}$ the reduced density matrices ${\rho}_1$ and ${\rho}_2$ are of the form

$$U=\left(\matrix{\cos{\sigma\over 2}&-\sin{\sigma\over 2}e^{-i\varphi}\cr
\sin{\sigma\over 2}e^{i\varphi}&\cos{\sigma\over 2}\cr}\right),\quad
V=\left(\matrix{\cos{\tau\over 2}&-\sin{\tau\over 2}e^{-i\epsilon}\cr
\sin{\tau\over 2}e^{i\epsilon}&\cos{\tau\over 2}\cr}\right).\eqno(49)$$

\noindent
(Recall that $\varphi =\arg z$ and $\epsilon =\arg \zeta$, Eq. (42) and a similar one with $z$ replaced by $\zeta$.)
Putting these matrices in the first of Eq. (11)
we obtain the relations 

$$\sin{\tau\over 2}e^{i\epsilon}={\sin{\sigma\over 2}\sin{\theta\over 2}\over {\cos{\Delta\over 2}}}{\vert w/z\vert\over {\sqrt{1+\vert w/z\vert ^2}}}e^{\arg{w/z}},\eqno(50)$$

\noindent
and

$$\cos{\tau\over 2}={\cos{\sigma\over 2}\cos{\theta\over 2}\over   {\cos{\Delta\over 2}}}+
{ \sin{\sigma\over 2}\sin{\theta\over 2}\over {\cos{\Delta\over 2}}}{1\over 
{\sqrt{1+\vert w/z\vert ^2}}}.\eqno(51)$$

\noindent
to be used later.
A quick check for the phase of (50) is given by  relation (32) giving
$\epsilon\equiv\arg\zeta =\arg {w/z}$ which is confirmed by Eq. (50) too.

Let $\vert{\Phi}_1\rangle \equiv\vert v\rangle$ where $\vert v\rangle$ is
the (40) separable state a representative of the states over $\pi(\vert v\rangle )$! We know that the point $\pi(\vert v\rangle )$ is the nearest one to $\pi(\vert u\rangle)$. 
Consider now the (37) unique horizontal geodesic 
passing through the antipodal separable states $\vert v\rangle$ and $\vert v^{\prime}\rangle$  where the state $\vert{\Phi}_2\rangle=\vert v^{\prime}\rangle$ is defined by Eqs. (38-39). 
This horizontal geodesic in $S^7$ is also passing through the state
$\vert u\rangle{\langle u\vert v\rangle\over {\vert\langle u\vert v\rangle\vert}}$ which is apart from a quaternionic phase is our entangled state we have started with.
Explicitly we have the relation

$$
\vert u\rangle{\langle u\vert v\rangle\over {\vert\langle u\vert v\rangle\vert}}=\cos{\Delta\over 2}\vert v\rangle+\sin{\Delta\over 2}\vert v^{\prime}\rangle.\eqno(52)$$

\noindent
Multiplying this equation from the right by the quaternionic phase
${\langle v\vert u\rangle\over {\vert\langle v\vert u\rangle\vert}}$,
and noticing that $\cos{\Delta\over 2}$ and $\sin{\Delta\over 2}$ are just the
square roots of the eigenvalues of the reduced density matrices
we obtain the form which looks like the quaternionic version of the Schmidt decomposition

$$\vert u\rangle =\sqrt{{\lambda}_+}\vert v\rangle Q 
+\sqrt{{\lambda}_-}\vert v^{\perp}\rangle P, \quad
Q\equiv
{\langle v\vert u\rangle\over {\vert\langle v\vert u\rangle\vert}},\eqno(53)$$

\noindent
where $P$ is the quaternionic phase transforming $\vert v^{\prime}\rangle$
to $\vert v^{\perp}\rangle$ having the (47) standard form.
In order to show that it is indeed the Schmidt decomposition
we have to show that the quaternionic phase $Q$ is somehow representing the other qubit in the Schmidt decomposition belonging to the fiber.
In other words we have to show that the vector $\vert {\bf t}\rangle$ of (46)
corresponds to $Q$.
Using the (53) definition for $Q$, the sections (31) and (40) with $q=1$ and relations (50-51) one readily obtains

$$Q=
{\cos{\sigma\over 2}\cos{\theta\over 2}\over   {\cos{\Delta\over 2}}}+
{ \sin{\sigma\over 2}\sin{\theta\over 2}\over {\cos{\Delta\over 2}}}{1+(w/z){\bf j}\over
{\sqrt{1+\vert w/z\vert ^2}}}= \cos{\tau\over 2}+\sin{\tau\over 2}e^{i\epsilon}{\bf j}.\eqno(54)$$

\noindent
Since according to Ref. [12] a generic state $\vert \psi _Q\rangle$ in the fiber is defined as
$Q\equiv\vert{\psi} _Q\rangle= c_0\vert 0\rangle _Q+c_1\vert 1\rangle _Q$ where the
orthogonal states $\vert 0\rangle _Q$ and $\vert 1\rangle _Q$  are related to the
choice $Q=1$ and $Q={\bf j}$ respectively, the correspondence between $Q$ and $\vert{\bf t}\rangle$  is established.
Alternatively the reader can check by calculating $\vert v\rangle Q$  and then identifying the parameters $a,b,c,d$ using (12) and (5) that this quaternionic spinor represents the separable state

$$\left(\cos{\sigma\over 2}\vert 0\rangle _1+\sin{\sigma\over 2}e^{i\varphi}\vert 1\rangle _1\right)\otimes \left(\cos{\tau\over 2}\vert 0\rangle _2+\sin{\tau\over 2}e^{i\epsilon}\vert 1\rangle _2\right)\equiv \vert {\bf v}\rangle\otimes\vert {\bf t}\rangle.\eqno(55)$$

Based on these considerations we have the following geometrical representation of the Schmidt decomposition for two qubits.
Chose first a special entangled state represented by a quaternionic spinor in the standard form (i.e. by putting $q=1$ in Eq. (31)) and call this  $\vert u\rangle$.
This state is represented by a point $\hat{u}\equiv\pi(\vert u\rangle)$ in ${\bf HP}^1\sim S^4$.
Find the point $ \hat{v}$ in the separable submanifold $S^2\subset S^4$ nearest to $\hat{u}$ connected by the unique geodesic segment.
(If $\vert u\rangle$ is not maximally entangled then we have a unique solution.)
As a next step parallel transport $\vert u\rangle$ along this geodesic 
with respect to the instanton connection to the fiber over $\hat{v}$.
Then one pair from the biorthogonal Schmidt states is recovered as the standard (i.e. of the (40) form) section $\vert v\rangle$ over
$\hat{v}$ and as the quaternionic phase $Q$ between this section and the state obtained by parallel transport.
The real expansion coefficient multiplying this pair forming the separable state
$\vert v\rangle Q$ is just $\cos{\Delta\over 2}$ where $\Delta$ is the geodesic distance between $\hat{u}$ and $\hat{v}$.
Repeat the same process for the antipodal point $\hat{v}^{\perp}$  of $\hat{v}$
in $S^4$ to obtain the the other pair of Schmidt states with expansion coefficient beeing the corresponding geodesic distance. 

If we use the Hopf fibration with the first qubit belonging to the base and the second to the fiber the gauge degree of freedom is manifested in the freedom
for the choice of the unitary matrix $V$ in Eq. (11). 
This corresponds to a different choice for the local base belonging to the second qubit.
In our representation for an {\it arbitrary} entangled state it means that we have a quaternionic spinor $\vert u\rangle$ this time with some fixed $q\neq 1$.
Of course this new choice will not change our geometric interpretation of the Schmidt decomposition.  
It is easy to see that the new state representing the first qubit in the Schmidt decomposition is the same, and the second is obtained from $Q$ by right multiplication  by $q$.
Hence we have proved that the Schmidt decomposition for two qubits
is amenable for a nice geometric interpretation in terms of the anholonomy of the canonical connection on the Hopf fibration.
Instantons were originally introduced as classical solutions of $SU(2)$  gauge  theories in Wick rotated space-time.
They also play an important role in quantum field theories describing tunnelling between different vacua.
It is amusing to find them here as the basic entities describing a fundamental aspect of quantum theory, two qubit entanglement.

\bigskip
\bigskip
\centerline{\bf VI. Geometric phases}
\bigskip

We have seen that the instanton connection on the Hopf fibration plays a vital role for the geometrical description of entanglement for two qubits.
A further interesting possibility to 
explore is to
look at the non-Abelian (an)holonomy of the instanton connection, and reinterpret the results in the language of entanglement.
For this purpose we have to somehow generate closed curves in the base
manifold $S^4$ and calculate the quaternionic phases picked up by an initial quaternionic spinor after completing a circuit.
Reinterpreting these spinors as entangled states the result of this non-Abelian  parallel transport
will be some final entangled state with very different form but {\it the same concurrence}.
For different loops we obtain different anholonomy matrices, that can serve as quantum gates. This process is called anholonomic quantum computation [14].

However, $S^4$ is sliced to submanifolds of fixed entanglement, and we know that
for separable states the anholonomy of the connection is Abelian.
Therefore interesting curves generating non-abelian anholonomy
are the ones not restricted to the $S^2$ submanifold with concurrence ${\cal C}=0$.
Moreover, a look at the (33) expression of $A$ we see that the states characterized by the condition $\vert z\vert =0$ describe another two-sphere $S^2\subset S^4$, with another monopole-like gauge-field on it. 
According to Ref. [12] states  with $\vert z\vert =0$ are the ones with trivial Schmidt decomposition.
For curves lying in the submanifold  of these states the holonomy is again Abelian.

Let us suppose then that we have a curve $C$ lying entirely in                  ${\cal U}\subset S^4$ not belonging to any of the submanifolds described above.
In this case we have to calculate the quaternionic phase $q[C]$ as given by formula

$$q[C]={\bf P}e^{-\oint_C{\cal A} }.\eqno(56)$$

\noindent
A more managable form for $q[C]$ can be given by dividing the loop $C$ into 
$N$ segments characterized by the $N$ points ${\xi}_0,{\xi}_1,\dots {\xi }_N={\xi}_0\in {\cal U}\subset S^4 $. These points represent a whole family of gauge equivalent entangled states belonging to the fiber. Let $P_n\equiv P({\xi}_n)=\vert u({\xi}_n)\rangle
\langle u({\xi}_n)\vert$ be the projectors represented by the quaternionic spinors $\vert u_n\rangle$ parametrized by the coordinates $\xi_n$. It is clear that $P_n$ is independent of the choice of representatives.
By using $\langle u({\xi}_{n+1})\vert u ({\xi}_n)\rangle \sim I-A({\xi}_n)+\dots$ one can show that [21]

$$
\vert u(\xi_0)\rangle{\bf P}e^{-\oint_C  A}=
\lim_{N\to \infty}P(\xi_N)P(\xi_{N-1})\dots P(\xi_1)\vert u(\xi_0)\rangle
.\eqno(57)$$

\noindent
Notice that Eq. (57) can also be understood in the following way.
Given a curve $C$ and its division by $N$ points we can approximate it by a geodesic polygon.
If we have an initial vector $\vert u_0\rangle$  over the point $\xi_0$ 
we can parallel translate this vector to the fiber over the next point $\xi_1$
obtaining the vector 
$\vert u_1\rangle{\langle u_1\vert u_0\rangle\over {\vert\langle u_1\vert u_0\rangle\vert}}$. After the next step we get the state
$\vert u_2\rangle{\langle u_2\vert u_1\rangle\langle u_1\vert u_0\rangle\over {\vert\langle u_2\vert u_1\rangle\langle u_1\vert u_0\rangle\vert}}$.
Iterating this process and then taking the limit $N\to\infty$
by virtue of $\vert \langle u_{n+1}\vert u_n\rangle\vert = 1-{1\over 2}dl^2+\dots$ we get the path ordered product of projectors verifying Eq. (57).

Hence in order to describe curves in $S^4$ we have to characterize a one parameter family of rank one quaternionic projectors $P(t)$. These are two-by-two quaternion-Hermitian matrices consisting of a single dyadic product of quaternionic spinors.  
It is easy to see that such projectors are of the form

$$P(t)={1\over 2}\left(I+{\Gamma}_{\mu}{\xi}_{\mu}(t)\right),\eqno(58)$$

\noindent
where ${\xi}_{\mu}\in S^4$ and the $2\times 2$ quaternion-Hermitian matrices
${\Gamma}_{\mu}, \mu=0,\dots 4$ are defined as

$${\Gamma}_0=\left(\matrix{1&0\cr 0&-1\cr}\right),\quad
\Gamma_1=\left(\matrix{0&1\cr 1&0\cr}\right),\quad
{\Gamma}_{2,3,4}=\left(\matrix{0&-{\bf i},{\bf j}, {\bf k}\cr
{\bf i},{\bf j},{\bf k}& 0\cr}\right).\eqno(59)$$

\noindent
In order to prove this note that $\{\Gamma_{\mu},\Gamma_{\nu}\}=2I{\delta}_{\mu\nu}$,
hence $P^2=P$.
One can also prove that we have

$$\langle u \vert\Gamma_{\mu}\vert u\rangle ={\xi}_{\mu}\quad{\rm where}\quad
\vert u\rangle =\left(\matrix{u_0\cr u_1\cr}\right)
\eqno(60)$$

\noindent
which is just another way of describing the (15) quaternionic Hopf fibration.
Using this we have $\langle u\vert P\vert u\rangle =1$, i.e.
$P=\vert u\rangle\langle u\vert$.
Using the (21) section one can obtain an explicit formula for $P$ in terms of $x$. By virtue of the (26) correspondence between ${\xi}$ and $x$ we get Eq. (58) as we have claimed.

We know that the anholonomy of the instanton connection representing the parallel transport of entangled states is described by Eq. (57) with 
$P(\xi)$ given by (58).
However, it is desirable to simplify (57) by restricting our attention to a subclass of curves of the form $P(t)=U(t)PU^{\dagger}(t)$, where $P$ is a fixed projector representing the initial entangled state in $S^4$. 
$U(t)$ is a one-parameter family of $2\times 2$ quaternion unitary matrices, i.e. $U(t)\in Sp(2)\simeq Spin(5)$ ($Spin(5)/{\bf Z}_2\sim SO(5)$). Such matrices are generated by quaternion skew-Hermitian
matrices $S$ in the form $U(t)=e^{tS}$, where $S$ can be expressed in terms of the generators of $Spin(5)$ as $S={\alpha}_{\mu\nu}S_{\mu\nu}$ with 
$S_{\mu\nu}={1\over 4}[\Gamma_{\mu},\Gamma_{\nu}]$ and ${\alpha}_{\mu\nu}=-{\alpha}_{\nu\mu}$ are real parameters.
For a closed loop we should have $P(2\pi)=P(0)$.

Now suppose we have an initial entangled state $\vert u\rangle\in S^7$
corresponding to the projector $P=\vert u\rangle\langle u\vert$ and a closed loop $P(t)= U(t)PU^{\dagger}(t)$ satisfying $P(2\pi)=P(0)=P$.
This loop is defined by some choice for the real parameters ${\alpha}_{\mu\nu}$.
Then we have the formula ( Proposition 7.4 of [22])

$$
\lim_{N\to \infty}P(\xi_N)P(\xi_{N-1})\dots P(\xi_1)P
=e^{{t}S}\left(\cos({t}\vert\vert PSP\vert\vert)P-
{\sin({t}\vert\vert PSP\vert\vert)\over {\vert\vert PSP\vert\vert}}PSP\right)
\eqno(61)$$
\noindent
where $P=P(\xi_0)$, and the quaternionic matrix norm is
${\vert\vert B\vert\vert}^2=Tr_{\bf H}(B^{\dagger}B)$.
Using this result we obtain our final formula

$$
\vert u\rangle{\bf P}e^{-\oint_C  A}=
e^{{t}S}\left(\cos({t}\vert\vert PSP\vert\vert)-
{\sin({t}\vert\vert PSP\vert\vert)\over {\vert\vert PSP\vert\vert}}PS\right)\vert u\rangle\eqno(62)$$

\noindent
where
$\vert u\rangle\equiv \vert u(\xi_0)\rangle$ is the quaternionic spinor representing our initial entangled state.

As an explicit example we take

$$S={1\over 2}\Gamma_1\left(\cos\kappa\Gamma_3-\sin\kappa\Gamma_4\right)={1\over 4}{\alpha}_{\mu\nu}[\Gamma_{\mu},\Gamma_{\nu}] \quad 0\leq \kappa\leq 2\pi\eqno(63)$$

\noindent
meaning the real parameters are chosen as ${\alpha}_{13}=\cos\kappa$
and ${\alpha}_{14}=-\sin\kappa$.
One can check that $4S^2=-I$ hence $U(t)=e^{tS}=\cos(t/2)+\sin(t/2)2S$. 
Using this we have $U(2\pi)=-I$ meaning $P(2\pi)=P(0)$ for any initial projector. Hence for changing values for $\kappa$ we obtain a parametrized family of  loops $C_{\kappa}$.

Let us chose the initial state to be the maximally entangled Bell-state

$$\vert u\rangle ={1\over{\sqrt{2}}}\left(\matrix{1\cr {\bf j}\cr}\right)\mapsto\vert\psi\rangle ={1\over {\sqrt{2}}}\left(\vert 0\rangle _1\otimes \vert 0\rangle _2
+\vert 1\rangle _1\otimes \vert 1\rangle_2\right).\eqno(64)$$

\noindent
Since this vector is the eigenvector of the matrix $\Gamma_3$ the initial projector is
$P=P(\xi_0)={1\over 2}(I+\Gamma_3)$. This means that the initial vector $\xi_0\in S^4$ for our loop $C_{\kappa}$ has components $\xi_{0{\mu}}=(0,0,0,1,0)$.
Straightforward calculation shows that

$$PSP=-{1\over 4}\sin\kappa\left(\matrix{{\bf k}&{\bf i}\cr {\bf i}&-{\bf k}\cr}\right),\quad \vert\vert PSP\vert\vert ={1\over 2}\sin\kappa.\eqno(65)$$

\noindent
Putting these results into Eq. (62) we get the result

$$
{1\over {\sqrt{2}}}\left(\matrix{1\cr{\bf j}\cr}\right){\bf P}e^{-\oint_C  A}=-\left(\matrix{\cos(\pi\sin\kappa)&{\bf i}\sin(\pi\sin\kappa)\cr
{\bf i}\sin(\pi\sin\kappa)&\cos(\pi\sin\kappa)\cr}\right){1\over {\sqrt{2}}}\left(\matrix{1\cr{\bf j}\cr}\right).\eqno(66)$$

\noindent
Hence after parallel transporting the Bell state $\vert u\rangle$ along $C$
it will pick up a quaternionic phase of the form

$$
q[C]={\bf P}e^{-\oint_C  A}=-\cos(\pi\sin\kappa)-\sin(\pi\sin\kappa){\bf k}.\eqno(67)$$

\noindent
From this equation and the (12) definition one can read off the complex
coefficients identifying the new entangled state.
These are $a=d=-\cos(\pi\sin\kappa)$ and $b=c=-i\sin(\pi\sin\kappa)$.
Since ${\cal C}=\vert ad-bc\vert =1$ we see that the anholonomy for our families of closed loops $C_{\kappa}$ is not changing the degree of entanglement.
This is not surprising since as the reader can check the $SU(2)\sim Sp(1)$ anholonomy transformations associated with an arbitrary loop $C$ belong to the local transformations
manipulating only the second qubit.

For the choice $\kappa ={\pi\over 6}$ we have 
$q[C]=-{\bf k}$. In this case the new state is 

$$\vert u\rangle q[C]=-{1\over {\sqrt{2}}}\left(\matrix{{\bf k}\cr {\bf i}\cr}\right)\mapsto \vert {\psi}^{\prime}\rangle =-{1\over {\sqrt{2}}}\left(
\vert 0\rangle _1\otimes \vert 1\rangle _2 +\vert 1\rangle _1 \otimes \vert 0\rangle _2\right)\eqno(68)$$

\noindent
hence for the loop $C_{\pi/6}$ we obtain up to a sign another Bell state.

We have not clarified the nature of our loops $C_{\kappa}$, $0\leq \kappa\leq 2\pi$ yet. 
In order to do this we write out explicitly $U(t)$ as

$$U(t)=\left(\matrix{\cos{t\over 2}+\sin{t\over 2}({\bf j}\cos\kappa-{\bf k}\sin\kappa)&0\cr 0&\cos{t\over 2}-\sin{t\over 2}({\bf j}\cos\kappa-{\bf k}\sin\kappa)\cr}\right),\eqno(69)$$

\noindent
and calculate the matrix

$$U(t)\left(\matrix{0&-{\bf j}\cr {\bf j}&0\cr}\right)U^{\dagger}(t)=\left(\matrix{{\xi}_0(t)&\overline{\xi}(t)\cr {\xi}(t)&-{\xi}_0(t)\cr}\right).\eqno(70)$$

\noindent
Straightforward calculation shows that the family of curves $C_{\kappa}$ in $S^4$ is given by

$${\xi}_{\mu}(t;\kappa)=\left(\matrix{0\cr\sin t\cos\kappa\cr 0\cr {\cos}^2{t\over 2}-{\sin}^2{t\over 2}\cos 2\kappa\cr {\sin}^2{t\over 2}\sin 2\kappa\cr}\right).\eqno(71)$$

\noindent
Noticing that ${\xi}_0=0$ we see that our family of loops
lies in the equator of $S^4$ which is a three-sphere $S^3$. 
Moreover, recalling Eq. (17) we see that $z(t;\kappa)=\sin t\cos\kappa$
and $w(t;\kappa)={\cos}^2{t\over 2}-{\sin}^2{t\over 2}e^{-2i\kappa}$.
Using ${\cal C}=\vert w\vert$ the concurrence as a function of  $t$ and $\kappa$ can be determined

$${\cal C}=\sqrt{{1\over 2}\left(1+\cos^2 t-\sin^2t\cos 2\kappa\right)}.\eqno(72)$$

\noindent
We see that the loops visit submanifolds of different entanglement, except for the case $\kappa ={\pi\over 2}$,  $z=0$ when the loop degenerates to the starting point. It can be understood as follows. Our $U(t)$ is an element of the 
subgroup $Spin(4)\subset Spin (5)\simeq Sp(2)$.
$Spin(4)/{\bf Z}_2\simeq SO(4)$ is the four dimensional rotation group.
The terms $\Gamma_1\Gamma_3$ and $\Gamma_1\Gamma_4$ in (63) generate $SO(4)$ rotations
in the $13$ and $14$ planes respectively. Clearly for $\kappa ={\pi\over 2}$ we have merely $14$ rotations not changing the ${\xi}_3=w=1$ constraint characterizing our initial Bell state. 
On the other hand for $\kappa =0$ we have $13$ rotations. In this case we have $z(t)=\sin t$
and $w(t)=\cos t$ hence ${\cal C}=\vert\cos t\vert$. This loop starting from the maximally entangled submanifold then crosses the separable surface at $t={\pi\over 2}$ meeting the maximally entangled surface again at $t=\pi$ and gets back to the separable surface at $t={3\pi\over 2}$ and then to the initial state at $t=2\pi$.
The anholonomy matrix for this curve from (66) is $-I$ reproducing the sign change of quaternionic spinors under a $2\pi$ rotation.
For the case $\kappa ={\pi\over 6}$ 
studied in Eq. (68) we get for the concurrence
${\cal C}={1\over 2}\sqrt{3+\cos^2t}$ along the loop.

Since $S^4\simeq SO(5)/SO(4)$ (i.e. $SO(5)$ acts transitively on $S^4$) we can also consider more general loops generated by $SO(5)$ rotations. Such loops are generated by an $S$ also containing terms of the form ${\alpha}_{0l}\Gamma_0\Gamma_k , k=1,2,3,4$.
There is also the possibility of considering loops generated by ordinary $SO(3)$ rotations. However, 
the structure of $SO(3)$ orbits  on $S^4$ is more complicated.
For the details of these orbits see Ref. [23], here we merely give
the (skew-Hermitian) $SO(3)$ generators needed to generate such loops 

$$J_1={1\over 2}\left(\sqrt{3}\Gamma_4\Gamma_0+\Gamma_3\Gamma_2+\Gamma_4\Gamma_1\right),$$
$$
J_2={1\over 2}\left(\sqrt{3}\Gamma_3\Gamma_0+\Gamma_1\Gamma_3+\Gamma_4\Gamma_2\right),\quad
J_3={1\over 2}\Gamma_3\Gamma_4+\Gamma_2\Gamma_1.\eqno(73)$$

\noindent
One can check that the usual relations $[J_j,J_k]=\epsilon_{jkm}J_m$ hold.
It is now clear that using an explicit form for $S$ and employing formula (61)
one can calculate the anholonomy of an arbitrary entangled state represented by a fixed projector $P$ for loops
regarded as $SO(n)$ orbits ($n=3,4,5$) of the form $P(t)=U(t)PU^{\dagger}(t)$.

It is now obvious that if we have a means of generating a prescribed set of closed loops on ${\bf HP}^1\simeq S^4$ through an entangled state of fixed concurrence 
then  we can associate to this loop an 
$SU(2)$ anholonomy matrix.
Moreover, this correspondence between loops and anholonomy matrices
can be useful in building one-qubit gates in the spirit of anholonomic quantum computation [14]. Due to the geometric nature of these gates, quantum information processing is expected to be fault tolerant.

With this possibility in sight the question arises:
how to generate loops on ${\bf HP}^1$?
The first possibility is the case of adiabatic evolution.
For this one can consider the parametrized family of
$2\times 2$ quaternion-Hermitian  Hamiltonians
$H(\xi)=\Gamma_{\mu}{\xi}_{\mu}$ reinterpreted as $4\times 4$ complex-Hermitian Hamiltonians with two Kramers degenerate doublets [24].
Note, however that unlike in Ref. [24] now we have to regard  $H(\xi)$ as a parametrized family of Hamiltonians coupling two qubits.  
In this picture we can view the entangled states as a parametrized family of eigenstates (eigensections) of these Hamiltonians.

In order to be more explicit we write $H(\xi)$ in the form

$$H(X)=X_{mn}J_mJ_n={\Gamma}_{\mu}\xi_{\mu}, \quad\mu=0,1,\dots 4,\quad m,n=1,2,3\eqno(74)$$

\noindent
where

$$X_{mn}={1\over \sqrt{3}}\left(\matrix{-\xi_1+{1\over {\sqrt{3}}}\xi_0&-\xi_2&-\xi_3\cr
-\xi_2&\xi_1+{1\over {\sqrt{3}}}\xi_0&\xi_4\cr
-\xi_3& \xi_4&-{2\over {\sqrt{3}}}\xi_0\cr}\right).\eqno(75)$$

\noindent
In this way we have made a mapping from $S^4$ to the space of real traceless symmetric $3\times 3$ matrices satisfying ${3\over 2}{\rm Tr}X^2=1$  i.e. to the space of unit quadrupoles.
$H(X)$ are precisely the quadrupole Hamiltonians studied by Avron et.al in Ref. [24]. 
However, now these Hamiltonians have a different interpretation.
We are not regarding these operators as describing a spin ${3\over 2}$ particle in an adiabatically changing quadrupole electric field with the underlying Hilbert space beeing ${\bf C}^4$.
We rather regard them as a parametrized set of coupling Hamiltonians for two qubits with Hilbert space ${\bf C}^2\times{\bf C}^2$.
This correspondence provides a nice formalism to
label entangled states with the eigenstates of quadrupole Hamiltonians. 
For example
separable states ($\xi_3=\xi_4=0$) are represented with a block diagonal quadrupole. Maximally entangled states are the ones with a complementary structure
for $X_{mn}$, i.e. $\xi_0=\xi_1=\xi_2=0$.
The Bell state of (64) is represented by the quadrupole
with the only nonvanishing components $X_{13}=X_{31}=-1/\sqrt{3}$.

Now we fix a unit quadrupole representing an entangled state. Changing the quadrupole components (i.e. the coupling between the two subsystems ) then we adiabatically trace out a loop $C$ in $S^4$.
Thanks to the adiabatic theorem [19]  the time dependent Schr\"odinger equation maps an initial eigenstate into an instantaneous eigenstate of the quadrupole Hamiltonian hence this state evolves along an open  curve in $S^7$ (the set of normalized states in our Hilbert space ${\cal H}\simeq {\bf C}^2\times{\bf C}^2$ ) whose shadow in $S^4$ is $C$.
The difference between the initial and final state in $S^7$ is an $SU(2)$ matrix (an $Sp(1)$ quaternionic phase) containing a dynamic and a geometric part.  
The geometric part in this case can be separated, it is precisely the anholonomy matrix corresponding to the operation of our quantum gate. 
In this way we have implemented quantum computation via quantum adiabatic evolution in the slowly changing environment of unit quadrupoles.

The other possibility for realizing closed curves on ${\bf HP}^1$ is
by the non-Abelian version of the Aharonov-Anandan phase [25]. 
In this case the assumption of adiabaticity is relaxed, by introducing 
cyclic solutions of the time dependent Schr\"odinger equation.
These are such solutions of the Schr\"odinger equation in $S^7$  whose shadow curves are closed in ${\bf HP}^1$. (In this case the solution curve is {\it not} an eigenstate of the instantaneous Hamiltonian.)  
In order to obtain an example of a cyclic evolution process in the quaternionic representation we have to find a $2\times 2$  quaternion unitary  matrix 
$U(t)$ and a $T\in {\bf R}$ for which we have

$$U(T)\vert u(0)\rangle =\vert u(0)\rangle p,\quad p\in Sp(1).\eqno(76)$$

\noindent
We can specify such a $U(t)$ by giving its quaternion skew-Hermitian generator $S$. A nice example of this kind is given by the choice

$$S(t)={1\over 4}[H(X(t)),H(\dot{X}(t))]={1\over 4}[H(\eta (t)), H(\dot{\eta}(t))]={1\over 4}[\Gamma_{\mu},\Gamma_{\nu}]{\eta_{\mu}(t)\dot{\eta}_{\nu}(t)},\eqno(77)$$

\noindent
for an ${\eta}_{\mu}(t)$ defining a closed loop ($\eta_{\mu}(T)={\eta}_{\mu}(0)$) in {\it another four sphere}
$S_{\eta}^4$.
Hence $U(t)$ is a $Sp(2)\simeq Spin (5)$ rotation of the form $U(t)=e^{S(t)}$ with ${\alpha}_{\mu\nu}(t)\equiv({\eta}_{\mu}(t)\dot{\eta}_{\nu}(t)-
{\eta}_{\nu}(t)\dot{\eta}_{\mu}(t))$.
It is important not to mix the four-sphere $S_{\eta}^4$ defining the parameters of the $Spin(5)$ rotation with the other four-sphere ${\bf HP}^1\simeq S^4$ the "space of states" for the entangled two qubits stratified into submanifolds of fixed entanglement.
We can of course identify the parameters of $S_{\eta}^4$ with the space of unit quadrupoles (this is reflected in  the first equality in (77)),
but then the relationship between the quadrupole components and the coordinates of ${\bf HP}^1$ is not canonical as was in the case of adiabatic evolution.

It is easy to show that 
the time evolution operator along the curve $C^{\prime}$ in $S^7$ is precisely the operator of parallel transport along $\pi(C^{\prime})=C\subset {\cal U}\subset {\bf HP}^1$  i.e. we have [18]

$${\bf P}e^{-\int_{C^{\prime}}S(t)dt}\vert u(0)\rangle =
\lim_{n\to\infty}P(T)P({n-1\over n}T)\dots P({2\over n}T)P({1\over n}T)\vert u(0)\rangle ,\eqno(78)$$

\noindent
where $P(t)$ is the projector belonging to the quadrupole operator
$H(\eta(t))=H(X(t))=\Gamma_{\mu}{\eta}_{\mu}(t)$ (the relationship between $X_{mn}$ and $\eta_{\mu}$ is of the same form as in (75)).
Note however, that unlike in the adiabatic case $\vert u(0)\rangle$ is {\it not} an eigenvector of the instantaneous Hamiltonian corresponding to $S(t)$.
The evolution is nonadiabatic and cyclic.
Since the dynamical phase [19] is  
just the integral of $\langle u(t)\vert S(t)\vert u(t)\rangle \equiv 0$ where
$\vert u(t)\rangle$ are the instantaneous eigenstates of $H(t)$,
the anholonomy of the evolution is purely geometric.
Hence we see that quadrupole-like Hamiltonians are capable of generating
closed curves via the standard Schr\"odinger type of evolution (both adiabatic and nonadiabatic) in the stratification manifold of two-qubits, enabling an implementation for anholonomic quantum computation.
The quantum gates obtained in this way are anholonomy transformations of the instanton connection our basic entity governing the entanglement properties of two qubits.

\bigskip
\bigskip
\centerline{\bf VII. Density matrices}
\bigskip

We have seen that the base space of our fibration is $S^4$ which is parametrized by the reduced density matrix $\rho_1$ of Eq. (7).
In this section we want to make this statement more precise.
As we see from Eq. (44) the space of density matrices is the three dimensional unit ball ${\bf B}^3$. The relationship between ${\bf B}^3$ and $S^4$ is clarified by Eq. (28) which shows that the line element on $S^4$ is a combination of a line element on ${\bf B}^3$ 
 of the form

$$4dl^2_{B}={d{\cal C}^2\over {1-{\cal C}^2}}+(1-{\cal C}^2)d\Omega,\eqno(79)$$

\noindent
and the line element of a circle  $S^1_{\cal C}$, $dl^2_{S^1_{\cal C}}={\cal C}^2d\chi^2$
with its radius parametrized by the value of the concurrence ${\cal C}$.
The line element in (79) is the one corresponding to the Bures metric [26] on the space of density matrices defined by

$$d_B(\rho,\omega)=\sqrt{2-2{\rm Tr}(\omega^{1/2}\rho\omega^{1/2})^{1/2}},\eqno(80)$$

\noindent
where $\rho$ and $\omega$ are density matrices.
Indeed, by restricting our attention to $2\times 2$ density matrices for the infinitesimal form of (79) we obtain the relation
[27]

$$dl^2_B={1\over 2}{\rm Tr}(d\rho d\rho)+d({\rm Det}\rho)^{1/2}d({\rm Det}\rho)^{1/2}.\eqno(81)$$
\noindent
Parametrizing $\rho$ as in the first expressions of (44) and (45) 
and recalling that ${\rm Det}\rho ={\cal C}^2/4$
we obtain (79).
Hence the line element on $S^4$ is of the form

$$dl^2=4dl^2_B+{\cal C}^2d\chi^2.\eqno(82)$$

The space of $2\times 2$ density matrices ${\cal D}_2$ is stratified into
the submanifold of rank one density matrices ${\cal D}_2(1)\simeq {\partial}{\bf B}^3\simeq S^2$, and the submanifold of rank two density matrices ${\cal D}_2(2)\simeq {\rm Int}{\bf B}^3$.
We realize that rank one density matrices correspond to separable states with ${\cal C}=0$, and rank two density matrices to the entangled  ones with ${\cal C}>0$.

Having clarified the  correspondence between reduced density matrices and  the base space, we now turn to total space 
$S^7$ of the Hopf bundle.
$S^7$ can be regarded as the space of purifications for the reduced density matrix ${\rho}_1$ . The structure of these purifications was studied in [28].
Here we use these results to give some new insight to the geometry of two qubit entanglement. 

Let us denote by ${\cal M}_2$ the space of complex $2\times 2$ matrices with elements ${\Lambda}\in {\cal M}_2$  satisfying
the constraint ${\rm Tr}\Lambda{\Lambda}^{\dagger}=1$.
It is clear that in our case $\Lambda ={1\over {\sqrt{2}}}C$ of Eq. (1), moreover due to normalization  ${\cal M}_2\simeq S^7$.
Then the mapping $f:\Lambda\in {\cal M}_2\mapsto \rho_1=\Lambda{\Lambda}^{\dagger}\in{\cal D}_2$ defines a stratification (see also Eq. (2)).
This stratification is the union of fibre bundles

$${\cal M}_2=\bigcup_k{\cal M}_2(k)\to {\cal D}_2=\bigcup_k{\cal D}_2(k),\quad k=1,2\eqno(83)$$

\noindent
where ${\cal M}_2(k)$ denotes the manifold of rank $k$ matrices in ${\cal M}_2$.

For the nonsingular case ${\rm Det}{\Lambda}={1\over 2}{\rm Det}C={1\over 2}w \neq 0$.
Write $w=\vert w\vert e^{i\chi}={\cal C}e^{i\chi}$.
Since the quantity ${\Lambda}{\Lambda}^{\dagger}$ is invariant with respect to right multiplication by an element of $U(2)$  ( the unitary $V$ of Eq. (11)) the mapping $f_2:{\cal M}_2(2)\to {\cal D}_2(2)$ is  a principal bundle with $U(2)$ fiber. 
There is a subbundle ${\cal B}_2$ of this bundle defined by the constraint $\chi\equiv 0$, i.e. ${\rm Det\Lambda}\in {\bf R}_+$.
As one can check ${\cal B}_2$ is an $SU(2)\simeq Sp(1)$ bundle over ${\rm Int}{\bf B}^3$.
This bundle will turn out to be important in the following considerations.

For the singular case it was proved in Ref. [28] that we have
${\cal M}_2(1)\simeq S^3\times_{U(1)}S^3$ hence we have an imbedding of the complex Hopf bundle into ${\cal M}_2(1)$. This is just the precise mathematical statement of the observation in [12] that in this case the quaternionic Hopf fibration can be "iterated" to include the complex Hopf fibration.
Moreover, in the light of this result it is not surprising that we have obtained in Section IV. the magnetic monopole connection (the canonical connection on the complex Hopf bundle) on this stratum.

An interesting topic to discuss here is the relationship between Uhlmann's connection for parallel transport for mixed states [29]
and our connection governing the entanglement properties of our qubits.
Uhlmann introduced "amplitudes" of density matrices. In our case these are 
just the matrices ${\Lambda_1}$ and ${\Lambda_2}$ purifying  two different nonsingular reduced density matrices $\rho$ and $\omega$. 
It is understood that $\rho$ and $\omega$ correspond to the reduced density matrices of two different entangled states with coordinates $z_{1,2}$ , $w_{1,2}$ and concurrences ${\cal C}_{1,2}\neq 0$. 
This  means that we have $\rho =\Lambda_1{\Lambda}_1^{\dagger}={1\over 2}C_1C_1^{\dagger}$ and
 $\omega =\Lambda_2{\Lambda}_2^{\dagger}={1\over 2}C_2C_2^{\dagger}$.
According to Uhlmann two such amplitudes are parallel iff

$$C_1^{\dagger}C_2=C_2^{\dagger}C_1>0.\eqno(84)$$

\noindent
It was also shown that on ${\cal M}_2(2)$ this parallel transport law is
implemented by a connection form ${\cal A}$  satisfying 

$$C^{\dagger}dC-dC^{\dagger}C=C^{\dagger}C{\cal A}+{\cal A}C^{\dagger}C.\eqno(85)$$

\noindent
Following Ref. [28] it is straightforward to check that Uhlmann's connection ${\cal A}$ on the subbundle ${\cal B}_2$ is just our $A$ instanton connection.
First we take the trace of (85)
then we use the identity $MN+NM={\rm Tr}(M)N+M{\rm Tr}(N)+({\rm Tr}(MN)-{\rm Tr}(M){\rm Tr}(N))$   
valid for our $2\times 2$ complex matrices
to get

$${\cal A}={1\over 2}\left(C^{\dagger}dC-dC^{\dagger} C\right)-{1\over 4}{\rm Tr}(C^{\dagger}dC-dC^{\dagger}C).\eqno(86)$$

\noindent
Here we have taken into account ${\rm Tr}(C^{\dagger}C)=2$ and
the restriction ${\rm Tr}({\cal A})=0$ which is valid on the subbundle ${\cal B}_2$ [28].
Now writing out $A={\rm Im}\langle u\vert du\rangle$ as a pure imaginary quaternion-valued  one form in the quaternionic notation
of (12) and  ${\cal A}$ as an $su(2)$-valued one form in the complex notation
for $C$ as given by (5), one can see that the two connections coincide.
It is important to stress once again that ${\cal A}=A$ only for 
entangled states lying in the subbundle with $\chi=\arg w=0$.
According to (82) for such states the line element of the metric on $S^4$ is just the Bures line element so the identification of the relevant connections is not surprising.
Since the connections for this class of entangled states coincide, so 
does the anholonomy properties of entangled states parallel transported in the subbundle  ${\cal B}_2$ of ${\cal M}_2(2)$.
Using the results of the previous section we can generate curves in the space of reduced density matrices and calculate the anholonomy of the entangled states regarded now as purifications lying in ${\cal B}_2$.
In this way in the context of two-qubit entanglement we managed to find a nice application for Uhlmann's parallel transport for density matrices. 

In the following we would like to make some comments on the
conformal structure of our metric on $S^4$.
Let us consider first the line element $dl^2_B$ associated with the Bures metric. Use the notation $R^2=\xi_0^2+\xi_1^2+\xi_2^2=1-{\cal C}^2$ and introduce the
new coordinate $-\infty \leq\beta\leq\infty$ via the relation
$R\equiv\tanh\beta$. Since $-1\leq R\leq 1$ this is a good parametrization for $R$.
Notice that $\beta$ behaves as the rapidity parameter in the special theory of relativity. For $\beta =0$ we obtain maximally entangled states and for $\beta=\pm \infty$ we get the separable states. 
It is straightforward to check that using this new parametrization the Bures line element takes the following form

$$4dl^2_B={1\over {\cosh^2\beta}}\left(d\beta^2+{\sinh^2\beta}d\Omega\right).\eqno(87)$$

\noindent
Since the line element $dl^2_{\bf H^3}=d\beta^2+\sinh^2\beta d\Omega$ is the line element on the upper sheet of the double sheeted hyperboloid ${\bf H}^3$
we can conclude that the Bures metric is conformally equivalent 
to the standard metric of hyperbolic geometry.
The hyperboloid ${\bf H}^3$ can be stereographically projected to the Poincar\'e ball ${\bf B}^3$ with the standard Poincar\'e metric on it.
Since the space of $2\times 2$ density matrices is just ${\bf B}^3$
the Bures metric is up to a conformal factor is just the Poincar\'e metric.
Hence the space of density matrices can be given a hyperbolic structure.
In this hyperbolic structure separable states are infinitely far away (i.e. on the boundary ${\partial}{\bf B}^3$) from  the maximally entangled ones.
This observation has already been made in a different context by Ungar in his
study of gyrovector spaces and the geometry of  $2\times 2$ density matrices [30].
We can take one step  more by realizing that the (28) line element for the four-sphere can also be given a conformal form after noticing that $(1-R^2)d\chi^2=
d\chi^2/\cosh\beta$ i.e. we have

$$dl^2={1\over {\cosh^2\beta}}\left(d\beta^2+\sinh^2\beta d\Omega +d\chi^2\right).\eqno(88)$$

\noindent
This means that conformally we have

$$S^4\simeq {\rm Int}{\bf B}^3\times S^1.\eqno(89)$$

\noindent
Notice that for this conformal equivalence we had to remove the boundary of ${\bf B}^3$ i.e. the separable states.
We can also understand Eq. (89) by looking at the standard line element on ${\bf R}^4$, $dl^2_{\bf R^4}=dx_1^2+dx_2^2+dx_3^2+dx_4^2$.
The $x_k, k=1,2,3,4$ are just the stereographically projected coordinates of (26).
Let $r$, and $\chi$ be polar coordinates on the $x_3-x_4$ plane.
According to (26) this plane is just the one corresponding to our complex coordinate
$w=\xi_3+{\bf i}\xi_4={\cal C}e^{i\chi}$.
Then we can write

$$dl^2_{\bf R^4}=r^2\left({dr^2+dx_1^2+dx_2^2\over {r^2}}+d\chi^2\right),\quad 0<r<\infty,\quad -\infty<x_1,x_2<\infty.\eqno(90)$$

\noindent
The metric $(dr^2+dx_1^2+dx_2^2)/r^2$ is the hyperbolic Poincar\'e metric of the upper half space ${\bf U}^3$ which can also be mapped to the  ${\bf B}^3$ Poincar\'e ball [31].
The metric  on ${\bf U}^3$ is singular at $r=0$ (separable states), hence
we have the identification ${\bf R}^4-{\bf R}^2\simeq {\bf U}^3\times S^1$.
After stereographic projection the left hand side is conformally equivalent to the four sphere $S^4$, and the right hand side by making use of the topological equivalence of ${\bf U}^3$ and ${\bf B}^3$ is just ${\bf B}^3\times S^1$. 

Closing this section as an interesting application of these ideas we calculate the geodesic distance between two entangled states 
$\vert \Psi\rangle$ and $\vert \Phi\rangle$ represented by the quaternionic spinors $\vert u\rangle$ and $\vert v\rangle$.
These spinors map to points $\xi_{\mu}$ and $\eta_{\nu}$ $\mu,\nu=0,1,\dots 4$ of $S^4$ via the projection of the Hopf fibration.
The first three components of the vectors $\xi_{\mu}$ and ${\eta}_{\nu}$ we denote by the vectors ${\bf u}$ and ${\bf v}$.
We relate as usual the remaning components to the concurrences 
 as $\xi_3+{\bf i}\xi_4={\cal C}_1e^{i\chi_1}$ and $\eta_3+{\bf i}\eta_4=   {\cal C}_2e^{i\chi_2}$.
We denote the reduced density matrices corresponding to $\vert u\rangle$ and $\vert v\rangle$ by $\varrho$ and $\omega$.
Then the transition probability related to the  (18) geodesic distance between our entangled states is given by the formula

$$\vert\langle u\vert v\rangle\vert^2={\rm Tr}_{\bf H}\left(P_uP_v\right)={1\over 2}(1+\xi_{\mu}\eta_{\mu})=
{1\over 2}\left(1+{\bf uv}+{\cal C}_1{\cal C}_2\cos(\chi_1-\chi_2)\right)
,\eqno(91)$$

\noindent
as can be checked by using the (58) form of our projectors and the Clifford algebra properties of the $\Gamma$ matrices.
For the subbundle ${\cal B}_2$ characterized by the constraint $\chi_1=\chi_2=0$
this expression according to Ref. [28] is just the Bures fidelity $[{\rm Tr}(\omega^{1/2}\rho\omega^{1/2})^{1/2}]^2$ occurring in the (80) Bures distance.
In terms of the rapidities $\beta_{\bf u}$ and $\beta_{\bf v}$
after introducing the Lorentz factors 
${\gamma}_{\bf u,v}=\cosh\beta_{\bf u,v}=1/\sqrt{1-\vert{\bf u,v}\vert^2}$ we can rewrite this expression for the distance of our entangled states as

$$\cos{\Delta_{uv}\over 2}=\vert\langle u\vert v\rangle\vert^2={1\over {2\gamma_{\bf u}\gamma_{\bf v}}}\left(\gamma_{\bf t}+\cos(\chi_1-\chi_2)\right),\quad
\gamma_{\bf t}\equiv \gamma_{\bf u}\gamma_{\bf v}(1+{\bf uv})\eqno(92)$$

\noindent
where $\gamma_{\bf t}$ is obtained by the addition law of velocities for ${\bf u}$ and ${\bf v}$ in special relativity
(the cosine theorem of hyperbolic geometry [30]).
As a special case for the subbundle ${\cal B}_2$  and for states with $\chi_1=\chi_2$ we obtain the formula of Ref. [32]
for the Bures fidelity amenable for a nice interpretation in hyperbolic trigonometry.
\bigskip
\bigskip
\centerline{\bf VIII. Conclusions}
\bigskip
In this paper we related the basic quantities of two-qubit entanglement to 
the geometrical structure of the quaternionic Hopf fibration.
The entangled state was represented by an element of the bundle space $S^7$.
One of our qubits was associated with the base ($S^4$) and the other with the fiber ($S^3$) of this fibration.
The nontriviality of the fibration i.e. the twisting of the bundle
have been connected with the entanglement of the qubits via the use of the canonical connection and the natural metric on $S^7$.
These quantities pull back via the use of sections to the base 
giving rise to the instanton gauge field with self-dual curvature, and the Mannoury-Fubini-Study metric.

  The base space can be startified to submanifolds of fixed entanglement.
Separable states occupy a two sphere $S^2$ , maximally entangled states a great circle $S^1$ of the equator of  ${\bf HP}^1\simeq S^4$.
The complement of separable states in the base can be conformally represented as $Int{\bf B}^3\times S^1$.
Here ${\rm Int}{\bf B}^3$ is the space of rank two reduced density matrices.
The measure of entanglement was defined as the length of the shortest geodesic
with respect to the Mannoury-Fubini-Study metric between the entangled state in question and the nearest separable state.
This nearest separabe state expressed in the form of the standard section
is one of the Schmidt states appearing in the Schmidt decomposition of the entangled state.
The other Schmidt state is reproduced from the quaternionic phase between this separable state and the one obtained by parallel transport along the shortest geodesic from our initial entangled state. 
The other pair of Schmidt states is obtained by a similar procedure
carried through by using the longer part of the geodesic, and the corresponding states.  

We examined the anholonomy properties realized by quaternionic geometric phases
for the two-qubit entangled states.
We have shown that for quadrupole-like Hamiltonians we can generate closed curves using the standard Schr\"odinger type of evolutions (both adiabatic and non-adiabatic) in the stratification manifold of two qubits.
These evolutions enable an implementation for anholonomic quantum computation.
For a specific family of curves we explicitly constructed the anholonomy transformations corresponding to a special class of quantum gates. 

By looking at the total space $S^7$ of our fibration as the space of 
purifications for our reduced density matrices we have started working out a dictionary between the non-Abelian $SU(2)$ geometric phase governing the entanglement of our qubits  and Uhlmann's law of parallel
transport  for purifications over the manifold of density matrices.
The space of such purifications forms a startification, i.e. a collection of fibre bundles. These bundles provide the natural geometrical setting for a deeper understanding of the results of Ref. [12].
We have shown how the Bures distance between reduced density matrices is related to the geodesic distance between entangled states.
We have also reformulated the  known relationship between the instanton connection and Uhlmann's connection from the viewpoint of two-qubit entanglement.
Based on results of Ungar et.al we made connection with the geometric data of the Hopf fibration, entanglement, and the hyperbolic geometry of the space of reduced density matrices.

It would be interesting to see whether the usefulness of quaternions  is merely a specific property of two qubit entanglement or it can be generalized for other
entangled quantum systems.
In the context of quaternionic quantum mechanics, or merely as a convenient representation we can consider the higher quaternionic Hopf fibrations $\pi: S^{4n+3}\to {\bf HP}^n$ with again an $Sp(1)$ fiber. 
The connection and  metric used here can be generalized and has already been discussed elsewhere [19]. 
In this case one can use $n+1$ component quaternionic vectors somehow representing entangled states. However, the structure of these bundles should be only capable
of incorporating the twisting of merely one qubit (belonging to the fiber) with respect to some other (possibly also entangled) state represented by the base degree of freedom.
The other possibility to utilize the nontrivial fibre bundle structure as
a geometric representation for quantum entanglement is to consider spin bundles , based on higher dimensional Clifford algebras. 
In particular the $2^n$ dimensional spin bundles over $2n$ dimensional
spheres seems to be promising. For the $n=2$ case we get essentially the Hopf fibration. For the general case the fibre  is $SU(2^{n-1})$,
and the tensor product structure of higher dimensional $\Gamma$ matrices hints
a possible use for understanding a special subclass of entangled systems
in the language of higher dimensional monopoles as connections on such bundles.
Apart from the fact whether these expectations are realized or not
we hope that we have convinced the reader that the quaternionic Hopf fibration   with its instanton connection is tailor made to unveil the basic structure  of two-qubit entanglement.

\bigskip
\bigskip
\centerline{\bf Acknowledgement}
\bigskip
Financial support from the Orsz\'agos Tudom\'anyos Kutat\'asi Alap (OTKA),
grant nos T032453 and T038191 is gratefully acknowledged.

\bigskip
\bigskip
\centerline{\bf REFERENCES}
\bigskip

\noindent
[1] J. S. Bell, {\it      Speakable and Unspeakable in Quantum Mechanics }
Cambridge University Press, 1987.

\noindent
[2] D. Deutsch 1989 Proc. Roy. Soc. London {\bf 400} 97, {\bf A425} 73

\noindent
[3] C. H. Bennett, G. Brassard, C. Crepeau, R. Jozsa, A. Peres and W. K. Wooters 1993 Phys. Rev. Lett. {\bf 70} 1895-1899

\noindent
[4] C. H. Bennett and  S. J. Wiesner 1992 Phys. Rev. Lett. {\bf 69} 2881-2884

\noindent
[5] A. K. Ekert 1991 Phys. Rev. Lett. {\bf 67} 661-663

\noindent
[6] W. K. Wooters 1998 Phys. Rev. Lett. {\bf 80} 2245-2248

\noindent
[7] S. Hill and W. K. Wooters 1997 Phys. Rev. Lett {\bf 78} 5022-5025

\noindent
[8] W. K. Wooters 2001 Quantum Information and Computation {\bf 1} 27-44

\noindent
[9] M. Ku\'s and K. ${\dot Z}$yczkowski 2001 Physical Review {\bf A63} 032307

\noindent
[10] I. Bengtsson, J. Br\"annlund and K. ${\dot Z}$yczkowski 2002 Int. J. Mod. Phys. {\bf A17} 4675-4695 

\noindent
[11] D. C. Brody and L. P. Hughston 2001 Journal of Geometry and Physics {\bf 38} 19-53

\noindent
[12] R. Mosseri and R. Dandoloff 2001 J. Phys. {\bf A34} 10243-10252

\noindent
[13] B. A. Bernevig and H. D. Chen 2003  e-print quant-ph/0302081

\noindent
[14] P. Zanardi and M. Rasetti 1999 Phys. Letters {\bf A264} 94-99

\noindent
[15] A. Peres, {\it Quantum Theory: Concepts and Methods}, Kluwer Academic, Dordrecht 1993

\noindent
[16] C. H. Bennett, H. J. Bernstein, S. Popescu, and B. Schumacher 1996
Phys. Rev. {\bf A53}, 2046-2052

\noindent
[17] G. Naber, {\it Topology, Geometry and Gauge Fields}, Springer-Verlag, New-York 2000 

\noindent
[18] P. L\'evay 1991 J. Math. Phys. {\bf 32} 2347-2357

\noindent
[19] A. Shapere and F. Wilczek (eds), {\it Geometric phases in physics}
World Scientific, 1989

\noindent
[20] N. Mukunda and R. Simon 1993 Ann. of Physics NY. {\bf 228}, 205-267

\noindent
[21] H. Kuratsuji and K Takada 1990 Modern Physics Letters {\bf A22}, 1765-1772

\noindent
[22] J. E. Avron, L. Sadun, J. Segert and B. Simon 1989 Commun. Math. Phys. {\bf 124} 595-627

\noindent
[23] L. Sadun and J. Segert 1989 J. Phys. {\bf A22} L111-L115

\noindent
[24] J. E. Avron, L. Sadun, J. Segert, and B. Simon 1988 Phys. Rev. Lett. {\bf 61} 1329-1332

\noindent
[25] J. Anandan 1988 Phys. Lett. {\bf 133A} 171, P. L\'evay 1990, Phys. Rev. {\bf A41} 2837-2840.

\noindent
[26] D. Bures 1969 Trans. Am. Math. Soc. {\bf 135} 199-212.

\noindent
[27] M. H\"ubner 1992 Phys. Lett. {\bf A163} 239-242.

\noindent
[28] J. Dittmann and G. Rudolph 1992 Journal of Geometry and Physics {\bf 10} 93-106.

\noindent
[29] A. Uhlmann 1986 Rep. Math. Phys. {\bf 24} 229-240, 1991 Lett. Math. Phys. {\bf 21} 229-236.

\noindent
[30] A. Ungar 2002 Foundations of Physics {\bf 32} 1671-1699

\noindent
[31] J. G. Ratcliffe, {\it Foundations of Hyperbolic Manifolds} Springer-Verlag
 1994.

\noindent
[32] J. Chen, L. Fu, A. A. Ungar and X. Zhao 2002 Physical Review {\bf A65} 024303(3).
\end